\documentclass[12pt]{JHEP3}
\usepackage{graphicx}
\usepackage{url}
\usepackage{amsmath}

\def\jpsi{{J/\psi}}
\def\psits{{\psi(2S)}}

\def\chico{{\chi_{c1}}}
\def\chict{{\chi_{c2}}}

\def\ss{{\bigl.^3\hspace{-1mm}S^{[1]}_1}}
\def\sps{{\bigl.^1\hspace{-1mm}S^{[8]}_0}}
\def\so{{\bigl.^3\hspace{-1mm}S^{[8]}_1}}
\def\pjs{{\bigl.^3\hspace{-1mm}P^{[1]}_J}}

\def\pj{{\bigl.^3\hspace{-1mm}P^{[8]}_J}}
\def\p0{{\bigl.^3\hspace{-1mm}P^{[8]}_0}}

\def\sa{{\bigl.^1\hspace{-1mm}S^{[8]}_0}}
\def\sb{{\bigl.^3\hspace{-1mm}S^{[8]}_1}}
\def\mo{\mathcal{O}}

\def\moss{{\langle\mathcal{O}^\jpsi(\bigl.^3\hspace{-1mm}S_1^{[1]})\rangle}}

\def\mopc{{\langle\mathcal{O}^\jpsi(\bigl.^3\hspace{-1mm}P_0^{[8]})\rangle}}
\def\mossts{{\langle\mathcal{O}^\psits(\bigl.^3\hspace{-1mm}S_1^{[1]})\rangle}}

\def\mosochij{{\langle\mathcal{O}^{\chi_{cJ}}(\bigl.^3\hspace{-1mm}S_1^{[8]})\rangle}}
\def\mosochiz{{\langle\mathcal{O}^{\chi_{c0}}(\bigl.^3\hspace{-1mm}S_1^{[8]})\rangle}}
\def\mopschij{{\langle\mathcal{O}^{\chi_{cJ}}(\bigl.^3\hspace{-1mm}P_J^{[1]})\rangle}}
\def\mopschiz{{\langle\mathcal{O}^{\chi_{c0}}(\bigl.^3\hspace{-1mm}P_0^{[1]})\rangle}}
\def\mopajpsit{{\langle\mathcal{O}^{\jpsi(\psits)}(\bigl.^1\hspace{-1mm}S_0^{[8]})\rangle}}
\def\mopbjpsit{{\langle\mathcal{O}^{\jpsi(\psits)}(\bigl.^3\hspace{-1mm}S_1^{[8]})\rangle}}
\def\mopcjpsit{{\langle\mathcal{O}^{\jpsi(\psits)}(\bigl.^3\hspace{-1mm}P_0^{[8]})\rangle}}
\def\to{\rightarrow}

\def\Majpsi{M_{0,r_0}^{\jpsi}}
\def\Mbjpsi{M_{1,r_1}^{\jpsi}}
\def\Mapsits{M_{0,r_0}^{\psits}}
\def\Mbpsits{M_{1,r_1}^{\psits}}

\def\Majpsit{M_{0,r_0}^{\jpsi(\psits)}}
\def\Mbjpsit{M_{1,r_1}^{\jpsi(\psits)}}
\def\bqa{\begin{eqnarray}}
\def\eqa{\end{eqnarray}}
\def\bc{\begin{center}}
\def\bc{\end{center}}

\def\be{\begin{equation}}
\def\ee{\end{equation}}
\def\bea{\begin{eqnarray}}
\def\eea{\end{eqnarray}}

\def\gev{\mathrm{~GeV}}

\def\th{\theta}

\normalbaselines

\title{Yields and polarizations of prompt $\jpsi$ and $\psits$ production in hadronic collisions}

\author{Hua-Sheng Shao$^{a}$\footnote{Present address: {\it PH Department, TH Unit, CERN, CH-1211, Geneva 23, Switzerland}.},Hao Han$^{a}$,Yan-Qing Ma$^{b,c}$,Ce Meng$^{a}$, Yu-Jie Zhang$^{d}$, Kuang-Ta Chao$^{a,c,e}$\\
$^a$School of Physics and State Key Laboratory of Nuclear
Physics and Technology, Peking University,
 Beijing 100871, China\\
$^b$Maryland Center for Fundamental Physics, University of Maryland, College Park, Maryland 20742, USA\\
$^c$Center for High Energy Physics, Peking University, Beijing
100871, China\\
$^d$Key Laboratory of Micro-nano Measurement-Manipulation and Physics (Ministry of Education) and School of Physics, Beihang University, Beijing 100191, China\\
$^e$Collaborative Innovation Center of Quantum Matter, Beijing 100871, China\\
E-mail: \email{huasheng.shao@cern.ch,hao.han@pku.edu.cn,yqma@umd.edu,\\
mengce75@pku.edu.cn,nophy0@gmail.com,ktchao@pku.edu.cn}}

\abstract{
We give predictions of 
$J/\psi$ and $\psi(2S)$ yields and polarizations in prompt production at hadron colliders
based on non-relativistic QCD factorization formula. We calculate short-distance coefficients of all important color-octet
intermediate channels as well as color-singlet channels up to
$\mathcal{O}(\alpha_S^4)$, i.e. next-to-leading order in $\alpha_S$. For prompt
$\jpsi$ production, we also take into account feeddown contributions from $\chi_{cJ}$(J=0,1,2) and $\psits$
decays. Color-singlet long-distance matrix elements (LDMEs) are estimated by using potential model, and color-octet LDMEs are extracted by fitting the Tevatron yield data only. 
The predictions are satisfactory for both yields and polarizations of prompt $\jpsi$ and prompt $\psits$
production at the Tevatron and the LHC. In particular, we find our predictions for polarizations of prompt $\jpsi$ production have only a little difference from our previous predictions for polarizations of direct $\jpsi$ production.

}


\begin{document}

\section{Introduction}
Heavy quarkonium physics provides an ideal laboratory to study QCD at
the interplay between perturbative and non-perturbative domains. Because of the large samples of $\jpsi$ and $\psits$ accumulated at the LHC, it is an opportune moment to study the quarkonium production mechanism. To this end, understanding the polarization of produced quarkonium is an attractive and  important issue~\cite{Brambilla:2010cs,Lansberg:2006dh}.

At the Tevatron, the polarization observable $\lambda_{\th}$ (or
$\alpha$) for both of prompt $\jpsi$ and prompt $\psits$ in their helicity
frame measured by CDF Collaboration~\cite{Affolder:2000nn} are close
to $0$, which are in contradiction with the prediction of transverse polarization at the leading order (LO) in non-relativistic QCD (NRQCD)~\cite{Bodwin:1994jh}.
In the past a few years,
three groups~\cite{Butenschoen:2012px,Chao:2012iv,Gong:2012ug}
reported their independent analyzes of $\jpsi$ polarizations at the
next-to-leading order (NLO) level in $\alpha_S$. Although the short-distance coefficients (SDCs) are consistent with each other, the three groups give three different versions for the polarization prediction because different treatments are used in the extractions of
non-perturbative color-octet (CO) long-distance matrix elements
(LDMEs). Specifically, both ref.~\cite{Butenschoen:2012px} and ref.~\cite{Gong:2012ug} claim that the NLO NRQCD will necessarily give a transversely polarized prediction for prompt $\jpsi$ production; while the work by some of the present authors~\cite{Chao:2012iv} give a possible explanation for the $\jpsi$ polarization issue by finding that the transversely polarized contributions from $\so$ and $\pj$ channels cancel each other. A similar cancelation between $\so$ and $\pj$ channels for yield was also found earlier in ref.~\cite{Ma:2010yw}. The consequence of these cancelations is that the $\sps$ channel will dominate, which results in an unpolarized prediction. Note that, the crucial point to get these cancelations is the introduction of a relatively large $p_T$ cutoff for data in the lower $p_T$ region.

 In the high $p_T$ region, however, large logarithms like $\ln(p_T^2/m_c^2)$ may ruin the convergence of perturbative expansion, thus resummation of these large logarithmic terms are needed. This can be done by using DGLAP evolution equations to resum terms in the leading power in $1/p_T$ expansion, and using double parton evolution equations derived in ref.~\cite{Kang:2014tta} to resum terms in the next-to-leading power in $1/p_T$ expansion. The first goal is achieved recently~\cite{Bodwin:2014gia}. By combining the NLO NRQCD result with the leading power resummation, authors in ref.~\cite{Bodwin:2014gia} find that contributions from $\so$ and $\pj$ channels should be almost canceled with each other and the produced $\jpsi$ is almost unpolarized, which is similar to our conclusion in ref.~\cite{Chao:2012iv}. This is encouraging because it implies that the qualitative results in the NLO NRQCD calculation are not changed by resummation.

Based on the NLO NRQCD calculation, a data-driven method is employed in ref.~\cite{Faccioli:2014cqa} to fit CO LDMEs. By investigating the behaviour of $\chi^2/d.o.f.$ for different $p_T$ cutoff, the authors push the $p_T$ cutoff for $\psits$ to even larger values, say, about $12\gev$. Then they found that the $\psits$ production is dominated by the $\sps$ channel and the polarization data of $\psits$ production can be explained, which is similar to the explanation of $\jpsi$ polarization in refs.~\cite{Chao:2012iv,Bodwin:2014gia}. Therefore, it seems possible that the polarizations of $\jpsi$ and $\psits$ can be explained in a unified way.

However, in ref.~\cite{Chao:2012iv}, as well as in refs.~\cite{Butenschoen:2012px,Bodwin:2014gia}, only the direct $\jpsi$ production contribution is considered. An estimation of the
impact of feeddown contributions to $\jpsi$ polarization is given in ref.~\cite{Shao:2012fs}, where it was
pointed out that the feeddown contributions should not change the polarization result too much. Yet, to be precise, it should be better to include the feeddown contributions rigorously since they may contribute a substantial amount of prompt $\jpsi$ production. Hence, the purpose of the present article is to do a comprehensive analysis for prompt $\jpsi$ production by including
the feeddown contributions from $\chi_{cJ}$ and
$\psits$ decays. Meantime, we also give predictions of
yields and polarizations for prompt $\psits$.


The remaining context is organized as follows. We first fix our
strategy for estimating the LDMEs in section \ref{sec:2}, and
then give our predictions for the yields and polarizations of $\psits$
and $\jpsi$ in the next two sections. A summary will be given in the last section.

\section{Strategy for estimating LDMEs\label{sec:2}}
\subsection{General setup}

Before going ahead, we first list some details that are used in this article.
The helicity-summed yields are calculated following the way
mentioned in refs.~\cite{Ma:2010yw,Ma:2010jj,Ma:2010vd}, while the
method of the polarisation is described in
refs.~\cite{Chao:2012iv,Shao:2012iz,Shao:2014fca}.

Cross section for a
quarkonium $\mathcal{Q}$ production in $pp$ collision can be
expressed as~\cite{Bodwin:1994jh}
\bqa
\sigma(pp\to\mathcal{Q}+X)&=&\sum_{n}{\hat{\sigma}(pp\to
Q\bar{Q}[n]+X)}\times\langle\mathcal{O}^{\mathcal{Q}}(n)\rangle,
\eqa where $\hat{\sigma}(pp\to Q\bar{Q}[n]+X)$ are SDCs for producing a heavy quark pair $Q\bar{Q}$ with the
quantum number $n$, and $\langle\mathcal{O}^{\mathcal{Q}}(n)\rangle$
is a LDME for $\mathcal{Q}$. SDCs can be computed in perturbative QCD as
\bqa \hat{\sigma}(pp\to
Q\bar{Q}[n]+X)&=&\sum_{a,b}{\int{dx_1dx_2d\rm{LIPS}
\textit{f}_{a/p}(x_1)\textit{f}_{b/p}(x_2)}}\nonumber\\&\times&|M(ab\to
Q\bar{Q}[n]+X)|^2,\eqa where the symbols $a$ and $b$ represent all
possible partons, $x_1$ and $x_2$ are light-cone momentum fractions, $d\rm{LIPS}$ is the
lorentz-invariant phase space measure, and
$\textit{f}_{a/p}(x_1)$ and $\textit{f}_{b/p}(x_2)$ are parton
distribution functions (PDFs) for partons $a$ and $b$ in the initial
colliding protons.

In this article, we have included all important $c\bar{c}$  Fock
states, $\ss,\sps,\sb$ and $\pj$ for $\jpsi$ and $\psits$,
$\sb$ and $\pjs$ for $\chi_{cJ}$. All corresponding SDCs are calculated up to
$\mo(\alpha_S^4)$, i.e. NLO in $\alpha_S$. We use
CTEQ6M~\cite{Pumplin:2002vw} as our default PDF. The mass of charm
quark is fixed to be $m_c=1.5 \rm{GeV}$, and an analysis of uncertainties from choosing charm quark mass can be found in ref.~\cite{Ma:2010yw}.
The renormalization and factorization scales are
$\mu_R=\mu_F=\sqrt{(2m_c)^2+p_T^2}$, while the NRQCD scale is
$\mu_{\Lambda}=m_c$. Since cross sections of charmonia are
decreasing with high powers of their $p_T$, we should consider the
$p_T$ spectrum shiftting in the decay of
$\mathcal{Q}_1\to\mathcal{Q}_0+X$ approximately by
$p_T^{\mathcal{Q}_0}=\frac{M_{\mathcal{Q}_0}}{M_{\mathcal{Q}_1}}p_T^{\mathcal{Q}_1}$~\cite{Ma:2010yw},
where $M_{\mathcal{Q}_0}$ and $M_{\mathcal{Q}_1}$ are physical masses for
quarkonia $\mathcal{Q}_0$ and $\mathcal{Q}_1$ respectively. Masses of relevant charmonia in our article are shown in
Table~\ref{tab:mass}. Table~\ref{tab:br} gives the branching
ratios for various decay processes involved in
this article.

\begin{table}
\begin{center}
\begin{tabular}{{c}*{4}{c}}\hline\hline
$\jpsi$ & $\psits$ & $\chi_{c0}$ & $\chi_{c1}$ & $\chi_{c2}$\\\hline
$3.097$ & $3.686$ & $3.415$ & $3.511$ & $3.556$\\\hline\hline
\end{tabular}
\end{center}
\caption{\label{tab:mass}Physical masses (in unit of GeV) of
various charmonia~\cite{Nakamura:2010zzi}.}
\end{table}

\begin{table}
\begin{center}
\begin{tabular}{{c}*{1}{c}}\hline\hline
decay channel & branching ratio ($\times10^{-2}$)\\\hline $\jpsi\to
\mu^+\mu^-$ & $5.93$\\
$\psits\to\mu^+\mu^-$ & $0.75$\\
$\psits\to\jpsi+X$ & $57.4$\\
$\psits\to\jpsi\pi^+\pi^-$ & $34.0$\\
$\psits\to \chi_{c0}+\gamma$ & $9.84$\\
$\psits\to \chi_{c1}+\gamma$ & $9.3$\\
$\psits\to \chi_{c2}+\gamma$ & $8.76$\\
$\chi_{c0}\to \jpsi+\gamma$ & $1.28$\\
$\chi_{c1}\to \jpsi+\gamma$ & $36.0$\\
$\chi_{c2}\to \jpsi+\gamma$ & $20.0$
\\\hline\hline
\end{tabular}
\end{center}
\caption{\label{tab:br}Branching ratios of various decay
processes involved in this article~\cite{Nakamura:2010zzi}.}
\end{table}

\begin{table}[h]
\begin{center}
\begin{tabular}{|{c}*{4}{|c}|} \hline
$p_{T\rm{cut}}^{\psits}$(GeV) &  $\Mapsits(\times10^{-2}\rm{GeV}^3)$ & $\Mbpsits(\times10^{-2}\rm{GeV}^3)$ & $\chi^2/d.o.f$  \\
\hline $5$ &
$1.3754\pm0.118931$ & $0.159987\pm0.0117348$ & $37.2068/16=2.32542$\\
$6$ & $1.93677\pm0.17044$ & $0.128511\pm0.0135506$ & $14.0112/14=1.0008$\\
$7$ & $2.23162\pm0.23115$ & $0.109918\pm0.0155178$ & $7.21501/12=0.601251$\\
$8$ & $2.253154\pm0.301835$ & $0.100531\pm0.0175978$ & $5.46679/10=0.546679$\\
$9$ & $2.7258\pm 0.401123$ & $0.0932409\pm0.0201979$ & $4.92587/8=0.615734$\\
$10$ & $3.23067\pm0.58727$ & $0.0763209\pm0.0247166$ & $3.37617/6=0.562696$\\
$11$ & $3.81594\pm0.784395$ & $0.0585894\pm0.0293102$ & $2.10933/5=0.421866$\\
$12$ & $3.67631\pm1.00394$ & $0.0625013\pm0.0341653$ & $2.05968/4=0.514919$\\
$13$ & $3.48695\pm1.30212$ & $0.0673741\pm0.0402811$ & $2.00752/3=0.669175$\\
$14$ & $3.02071\pm1.7219$ & $0.0784274\pm0.0483324$ & $1.83628/2=0.918141$\\
$15$ & $1.04558\pm2.34914$ & $0.121791\pm0.0597233$ & $0.308538/1=0.308538$
 \\\hline
\end{tabular}
\end{center}
\caption{\label{tab:fitcdfpsi2s}The values of $\Mapsits$ and $\Mbpsits$ by fitting the CDF data~\cite{Aaltonen:2009dm} with different $p_{T\rm{cut}}$, where $r_0=3.9,r_1=-0.56$.}
\end{table}

The polarisation observable $\lambda_{\th}$ for $\jpsi$($\psits$) is
defined as ~\cite{Noman:1978eh,Beneke:1998re} \bqa
\lambda_{\th}=\frac{d\sigma_{11}-d\sigma_{00}}{d\sigma_{11}+d\sigma_{00}},
\eqa where $d\sigma_{ij}(i,j=0,\pm1)$ is the $ij$ component in the
spin density matrix formula for $\jpsi$($\psits$). The full spin
correlation of $\chi_{cJ}$'s spin density matrix element and
$\jpsi$'s spin density matrix element including E1, M2 and E3
transitions has been explored in eq.~(C4) of ref.~\cite{Shao:2012fs}.
We use the normalized M2 amplitude $a^{J=1}_2=-6.26\times10^{-2}$
for $\chi_{c1}\to\jpsi+\gamma$, and the normalized M2 and E3
amplitudes $a^{J=2}_2=-9.3\times10^{-2}$ and $a^{J=2}_3=0$ for
$\chi_{c2}\to\jpsi+\gamma$, which are measured by CLEO
collaboration~\cite{Artuso:2009aa}. From the eq.~(C4) of
ref.~\cite{Shao:2012fs}, we notice that the $\lambda_{\th}$ is
squared-amplitude dependent. Hence, these extra spin-flip effects due
to M2 and E3 transitions are negligible. We still keep it here since
no extra effort is needed.

\subsection{LDMEs estimation} \label{sec:ldmes}

Because of spin symmetry, LDMEs $\mosochij$ and $\mopschij$ for $\chi_{cJ}$ have the relation
\bqa \mosochij=(2J+1)\mosochiz,\\
\mopschij=(2J+1)\mopschiz.
\eqa
Color-singlet LDME $\mopschiz$ can be estimated by the derivation of wavefunction at origin $R'(0)$
via
\bqa
\mopschiz=2N_c\frac{3}{4\pi}|R'(0)|^2,
\eqa
where $|R'(0)|^2=0.075~\rm{GeV}^5$ is calculated
in ref.~\cite{Eichten:1995ch} by using potential model. The
remaining CO LDME $\mosochiz$ should be determined by fitting
experimental data. In ref.~\cite{Ma:2010vd}, we used $p_T$ spectrum of
$\sigma_{\chi_{c2}\to\jpsi\gamma}/\sigma_{\chi_{c1}\to\jpsi\gamma}$
measured by CDF~\cite{Abulencia:2007bra} in our fitting procedure, and we got
\bqa
\mosochiz=(2.2^{+0.48}_{-0.32})\times10^{-3}\rm{GeV}^3,
\eqa
which is consistent with later studies~\cite{Gong:2012ug,Shao:2014fca,Jia:2014jfa}.
Moreover, we want to emphasize that this value is insensitive to $p_T$ cutoff in our fit, especially when $p_T>7\rm{GeV}$.

Similarly, CS LDMEs for $\jpsi$ and $\psits$ can also be estimated by potential model~\cite{Eichten:1995ch},
\bqa
\moss=2N_c\frac{3}{4\pi}|R_{\jpsi}(0)|^2=1.16~\rm{GeV}^3,\\
\mossts=2N_c\frac{3}{4\pi}|R_{\psits}(0)|^2=0.76~\rm{GeV}^3,
\eqa
although their precise values are in fact irrelevant in our analysis because their corresponding SDCs are too small in our interested $p_T$ regime.
The determination of three unknown CO LDMEs
for $\jpsi(\psi(2S))$ is more complicated and involved. Based on our
previous studies~\cite{Chao:2012iv,Ma:2010yw,Ma:2010jj}, we
summarize the following facts:
\begin{itemize}
\item In the regime $p_T>4m_c$, the short-distance coefficient of P-wave CO Fock state $\pj$ can
be nicely decomposed into a linear combination of the short-distance
coefficients of $\sa$ and $\sb$, \bqa
d\hat{\sigma}(\pj)=r_0\frac{d\hat{\sigma}(\sa)}{m_c^2}+r_1\frac{d\hat{\sigma}(\sb)}{m_c^2}.\eqa
$r_0$ and $r_1$ changes slightly with rapidity interval but almost
not changes with the center-of-mass energy $\sqrt{S}$ (see table I in
ref.~\cite{Ma:2010jj}). This makes it difficult to extract three
independent CO LDMEs by fitting helicity-summed yields data at hadron
colliders. Instead, one is restricted to be able to extract two
linear combinations of three CO LDMEs within convincing precision.
They are denoted as\bqa
\Majpsit\equiv\mopajpsit+r_0\frac{\mopcjpsit}{m_c^2},\\
\Mbjpsit\equiv\mopbjpsit+r_1\frac{\mopcjpsit}{m_c^2}. \eqa
Because $d\hat{\sigma}(\sa)$ and $d\hat{\sigma}(\sb)$ have mainly $p_T^{-6}$ and $p_T^{-4}$ behaviour respectively, values of $\Majpsit$ and $\Mbjpsit$ can roughly indicate the relative importance of $p_T^{-6}$ and $p_T^{-4}$ components. Using the Tevatron yields
data~\cite{Acosta:2004yw,Aaltonen:2009dm} with $p_T>7\rm{GeV}$, they are extracted as in ref.~\cite{Ma:2010yw}
\bqa
\Majpsi=(7.4\pm1.9)\times10^{-2}\rm{GeV}^3,\\
\Mbjpsi=(0.05\pm0.02)\times10^{-2}\rm{GeV}^3,
\eqa
with $\chi^2/d.o.f=0.33$ for $\jpsi$, and
\bqa
\Mapsits=(2.0\pm0.6)\times10^{-2}\rm{GeV}^3, \label{eq:set1a}\\
\Mbpsits=(0.12\pm0.03)\times10^{-2}\rm{GeV}^3, \label{eq:set1b}
\eqa
with $\chi^2/d.o.f=0.56$ for $\psits$, with $r_0=3.9$ and $r_1=-0.56$. Inspired by the recent work~\cite{Faccioli:2014cqa}, we are also trying to see what happens if we enlarge the $p_T$ cutoff in our fit. With CDF data only~\cite{Acosta:2004yw,Aaltonen:2009dm}, we found values of $\Mapsits$ and $\Mbpsits$ can be alternated by enlarging the $p_T$ cutoff as shown in table~\ref{tab:fitcdfpsi2s}, while it is not the case for $\jpsi$.
When the cutoff is larger than $11\rm{GeV}$, we have relatively stable and minimal $\chi^2$ value for $\psi(2S)$. We thus obtained another set of CO LDMEs for $\psi(2S)$ by choosing cutoff as $p_T=11\gev$,
\bqa \label{eq:set2}
\Mapsits=(3.82\pm0.78)\times10^{-2}\rm{GeV}^3,\\
\Mbpsits=(0.059\pm0.029)\times10^{-2}\rm{GeV}^3.
\eqa
For simplification, we will call this set of CO LDMEs as ``set II" in the remaining context, while nothing will be labeled if we use the default one extracted from $p_T>7\rm{GeV}$ data in eqs.~\eqref{eq:set1a} and \eqref{eq:set1b}.
\item The short-distance coefficient\footnote{In this article, we only consider the helicity frame.} $d\hat{\sigma}_{11}(\pj)$ has
the similar decomposition but into $d\hat{\sigma}_{11}(\sa)$ and
$d\hat{\sigma}_{11}(\sb)$. The non-trivial thing is that coefficient of $d\hat{\sigma}_{11}(\sb)$ in
$d\hat{\sigma}_{11}(\pj)$ decomposition is quite close to $r_1$  in
$d\hat{\sigma}(\pj)$ decomposition~\cite{Chao:2012iv}. Hence, it
still does not help a lot to fix the three independent CO LDMEs by
including polarisation data~\cite{Chao:2012iv}. Moreover, the value of $\Mbjpsit$ almost
control the weight of transverse component. The unpolarized data
really require a (very) small $\Mbjpsit$.
\item We assume that all of the CO LDMEs are
positive~\cite{Chao:2012iv}, which is in contrast with those given
in refs.~\cite{Butenschoen:2012px,Gong:2012ug} (see also refs.~\cite{Butenschoen:2010rq,Butenschoen:2011yh,Butenschoen:2012qr}).\footnote{Although the authors in ref.~\cite{Gong:2012ug} used the same $p_T$ cut and included the feed-down contribution in prompt $J/\psi$ production, they tried to extract three independent CO LDMEs by including data in the forward-rapidity region. However, due to the correlation between the decompositions in the central and forward regions (see table I in ref.~\cite{Ma:2010jj}), the uncertainties in the extracted three CO LDMEs might be underestimated and they got negative CO LDMEs.} Since $r_1$ in forward
rapidity interval is smaller than that in central rapidity interval~\cite{Ma:2010jj},
a positive $\mopcjpsit$ would imply that $\lambda_\theta$ in forward rapidity
will be smaller than its value in the central
rapidity. We will see later that this conclusion is confirmed by LHC data. Further more, in a recent study of
$\jpsi+\gamma$ production~\cite{Li:2014ava}, the authors found that positivity of CO LDMEs are needed to guarantee a
physical cross section, while the sets of CO LDMEs in refs.~\cite{Butenschoen:2012px,Gong:2012ug} result in
unphysical negative cross section for $\jpsi+\gamma$ production at
hadron colliders. It also supports our assumption.
\end{itemize}
Based on these reasons, we are trying to use only Tevatron yield data as input to
give all yields and polarisation predictions for prompt
$\jpsi$ and $\psits$ production at hadron colliders. We use values of $\Majpsit$ and $\Mbjpsit$ in this section
and vary $0\leq\mopajpsit\leq\Majpsit$ to estimate the three CO
LDMEs. This variation and the errors in $\Majpsit$, $\Mbjpsit$ and
$\mosochiz$ will be considered as theoretical uncertainties.

\section{Prompt $\psits$ yields and polarizations\label{sec:3}}
In this section, we discuss the prompt $\psits$ yields and
polarisation at the Tevatron and the LHC. Experimentally, people can
reconstruct $\psits$ via $\psits\to\mu^+\mu^-$ or
$\psits\to\jpsi(\to\mu^+\mu^-)\pi^+\pi^-$. Unlike prompt $\jpsi$,
there is no significant feeddown contribution to prompt $\psits$
production.

\subsection{Yields}

\begin{figure}[!h]
\begin{center}
\hspace{0cm}\includegraphics[width=0.4\textwidth]{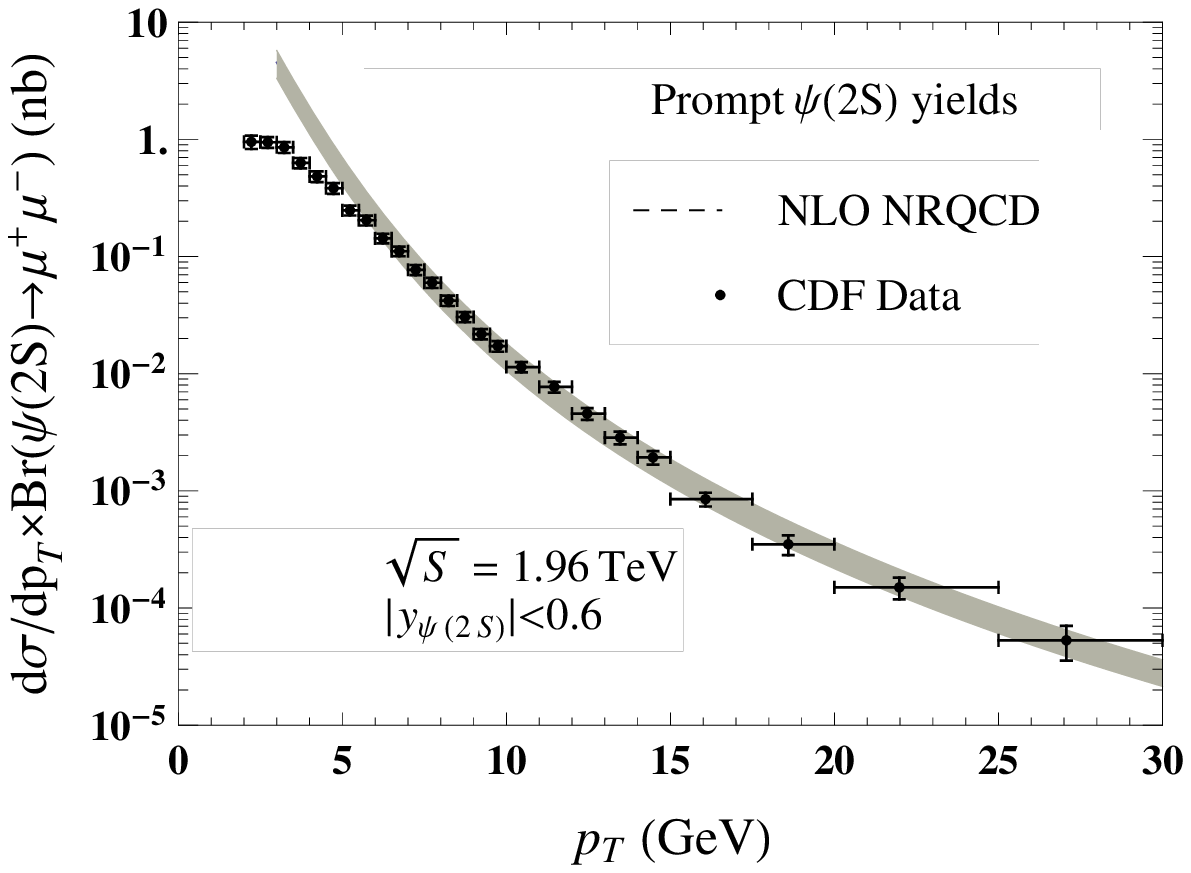}
\hspace{0cm}\includegraphics[width=0.4\textwidth]{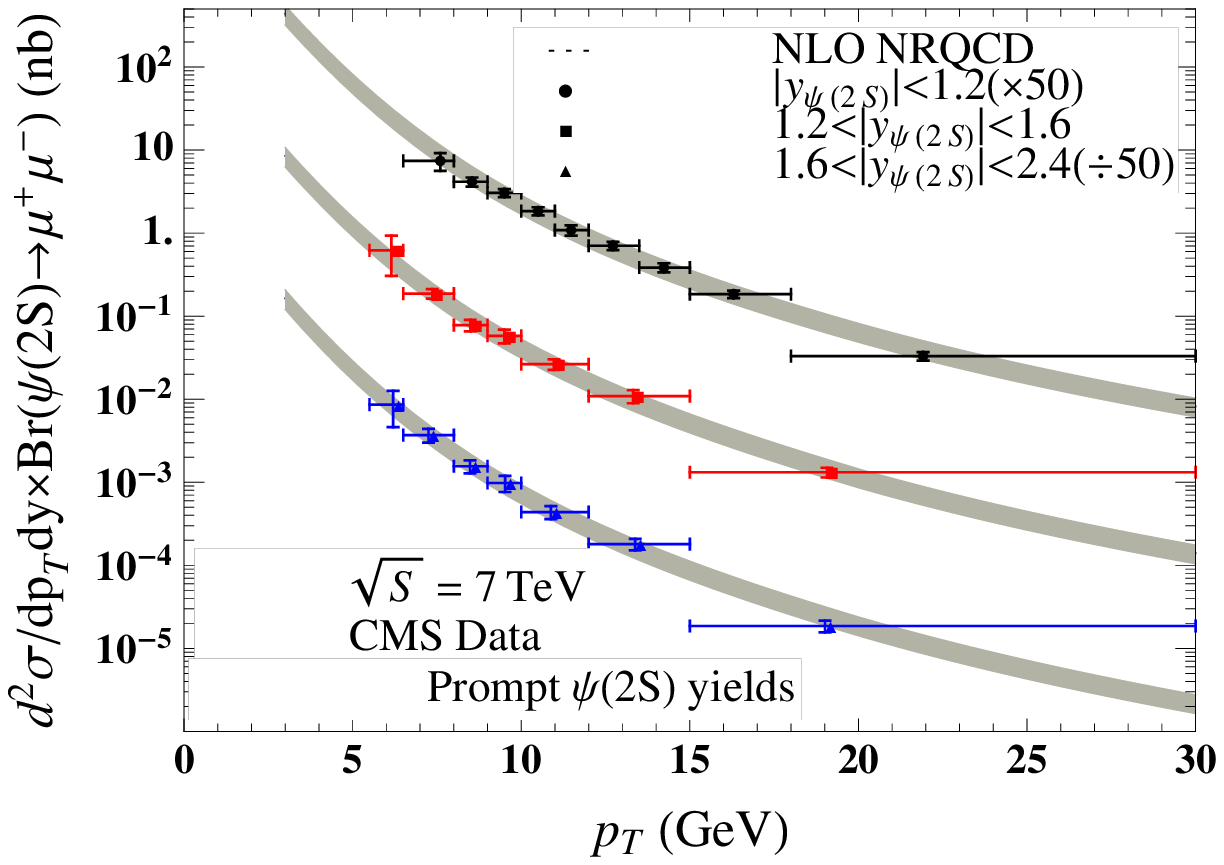}
\hspace{0cm}\includegraphics[width=0.4\textwidth]{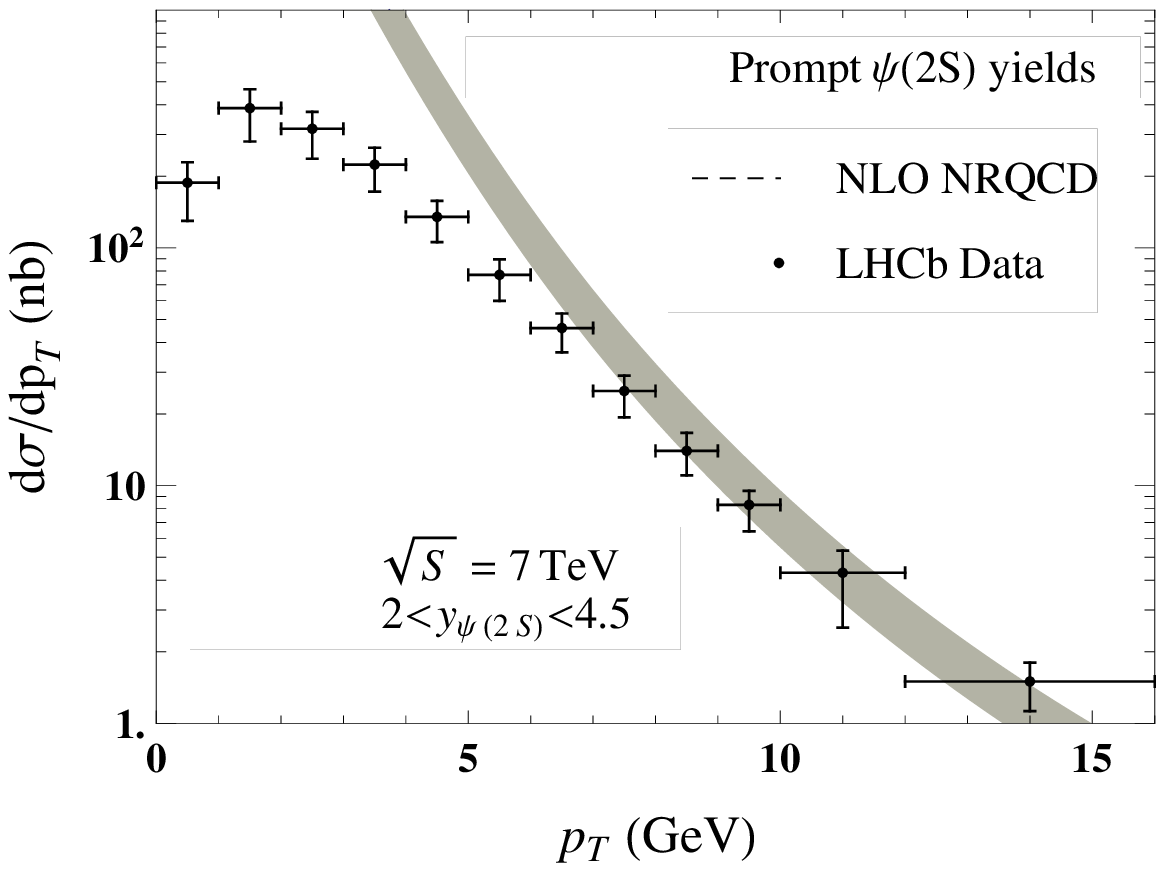}
\hspace{0cm}\includegraphics[width=0.4\textwidth]{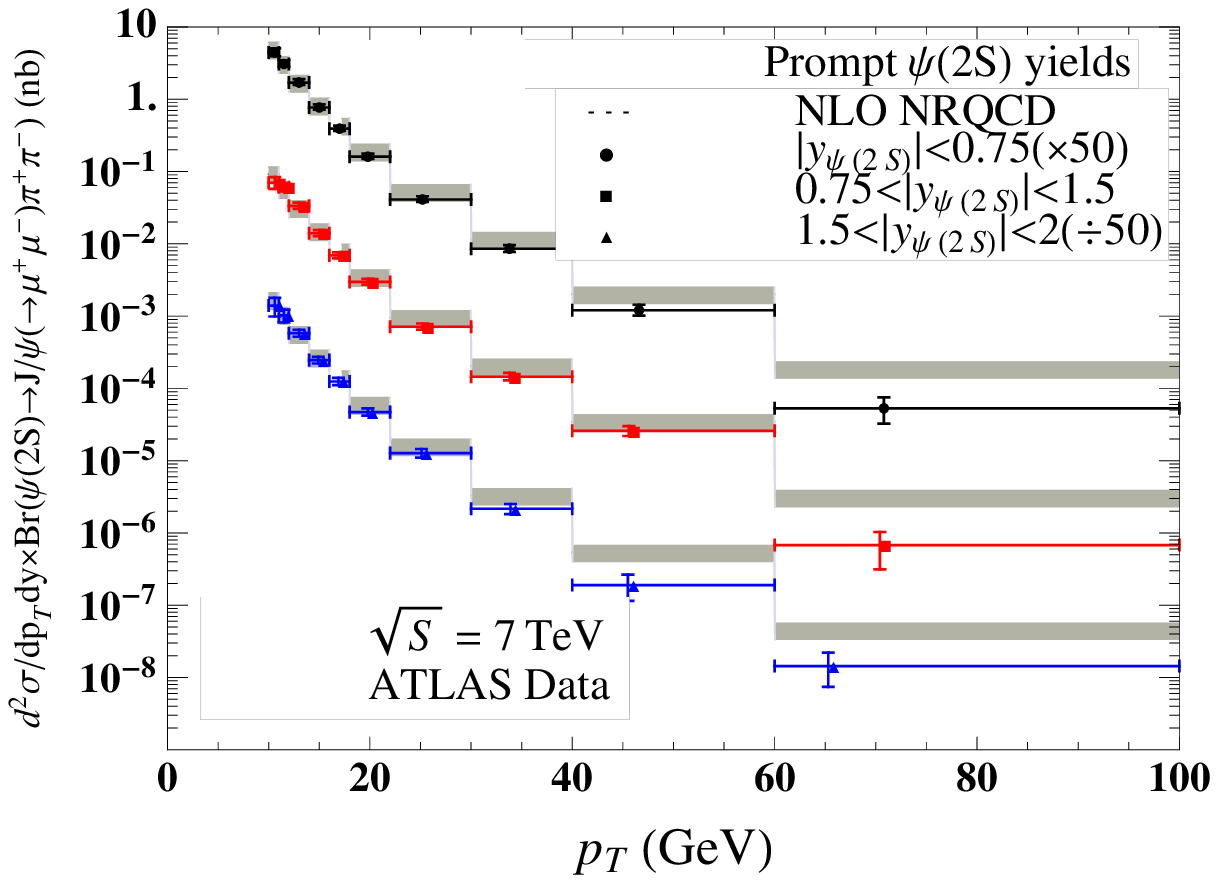}
\caption{\label{fig:psi2syields}Comparison of NLO NRQCD ( with the default set of CO LDMEs ) and
CDF~\cite{Aaltonen:2009dm},CMS~\cite{Chatrchyan:2011kc},
LHCb~\cite{Aaij:2012ag} and ATLAS~\cite{Aad:2014fpa} data for prompt $\psits$ yields.}
\end{center}
\end{figure}

\begin{figure}[!h]
\begin{center}
\hspace{0cm}\includegraphics[width=0.4\textwidth]{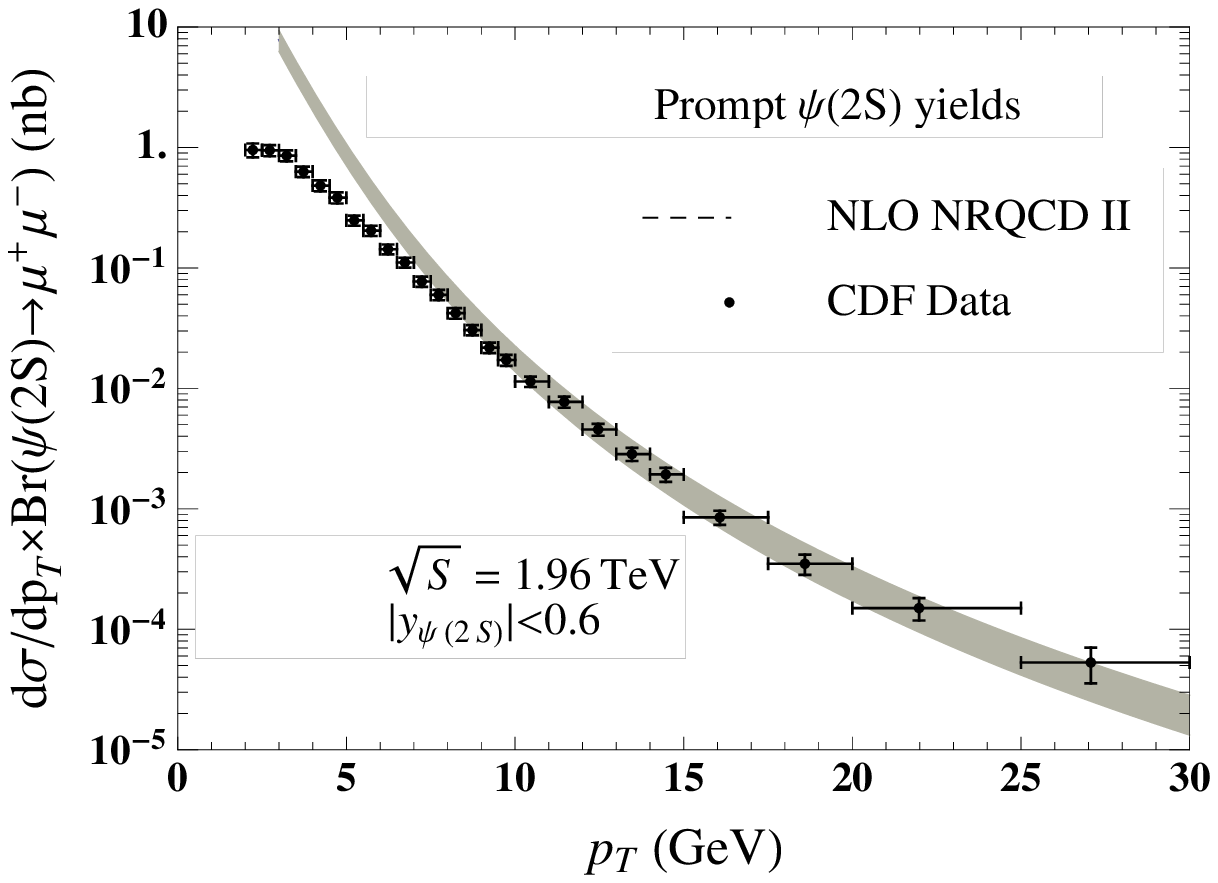}
\hspace{0cm}\includegraphics[width=0.4\textwidth]{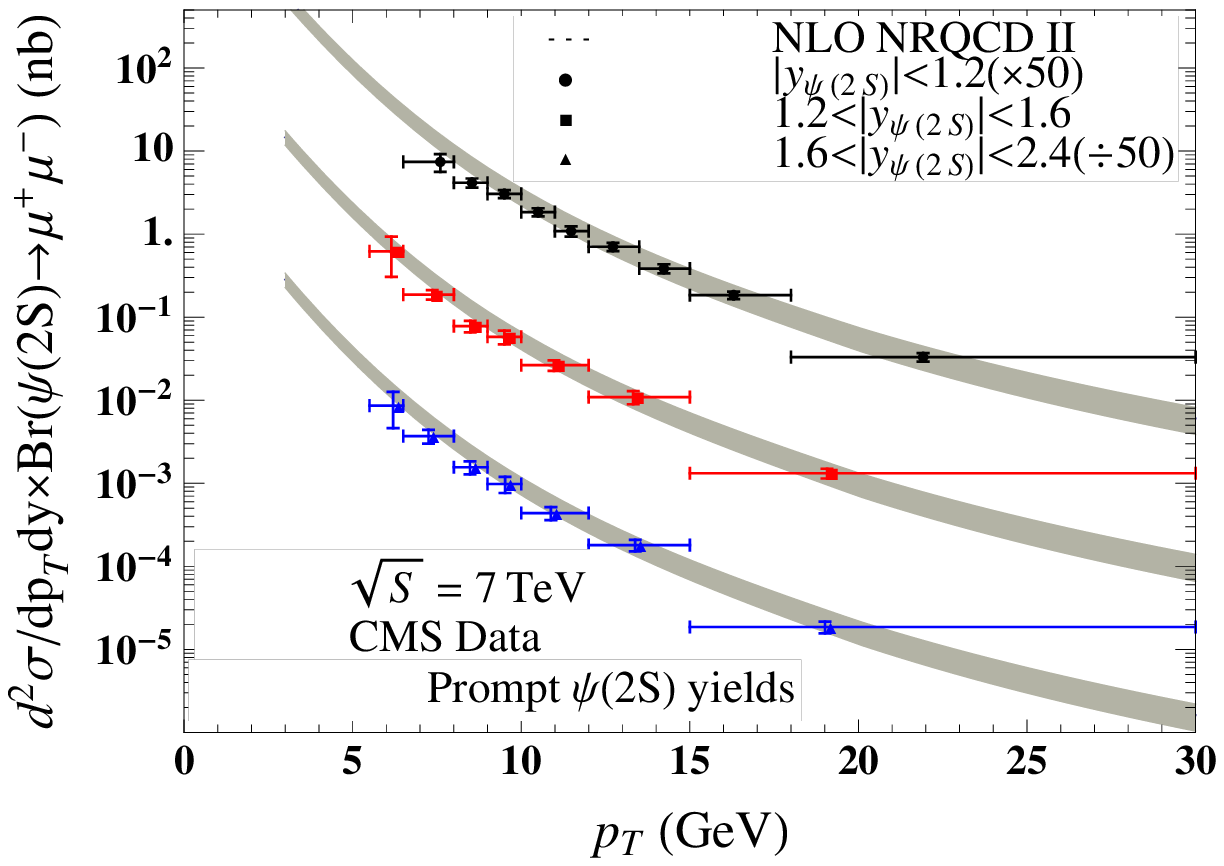}
\hspace{0cm}\includegraphics[width=0.4\textwidth]{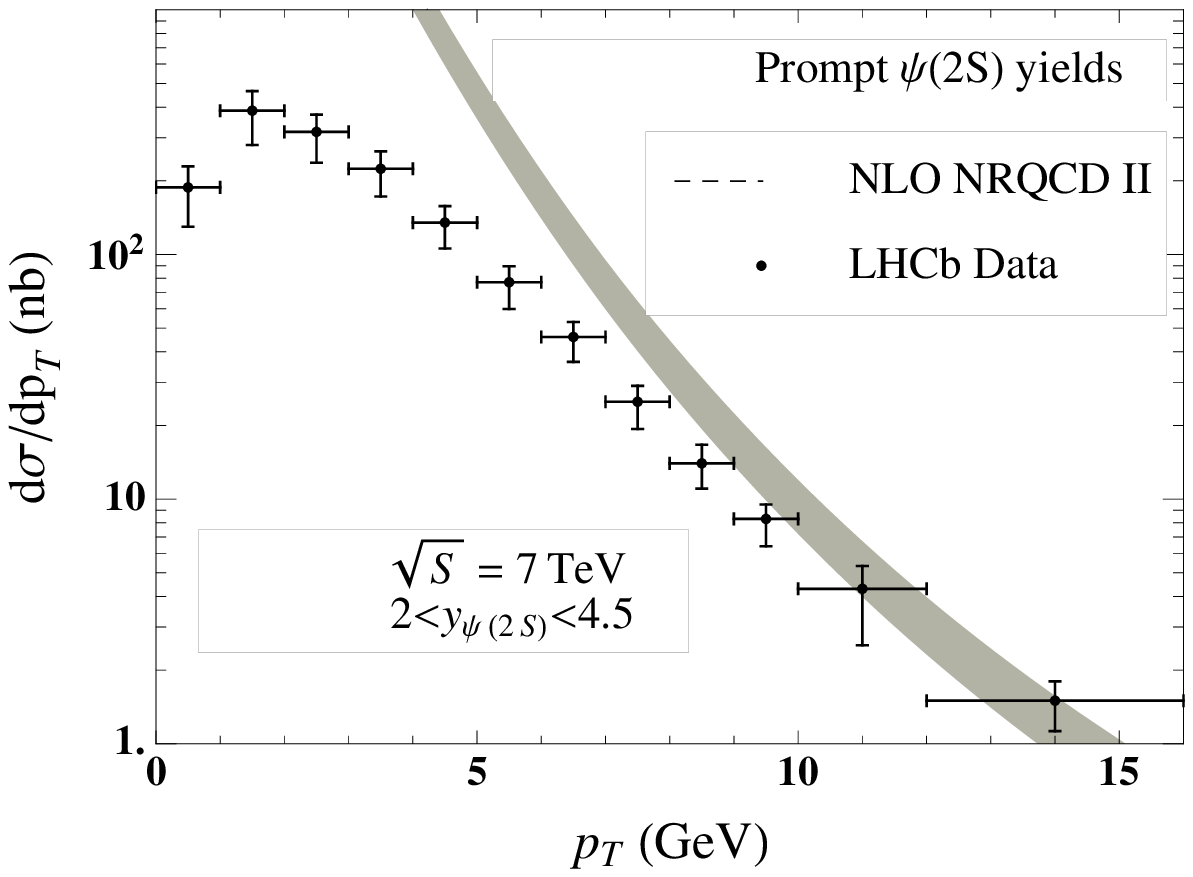}
\hspace{0cm}\includegraphics[width=0.4\textwidth]{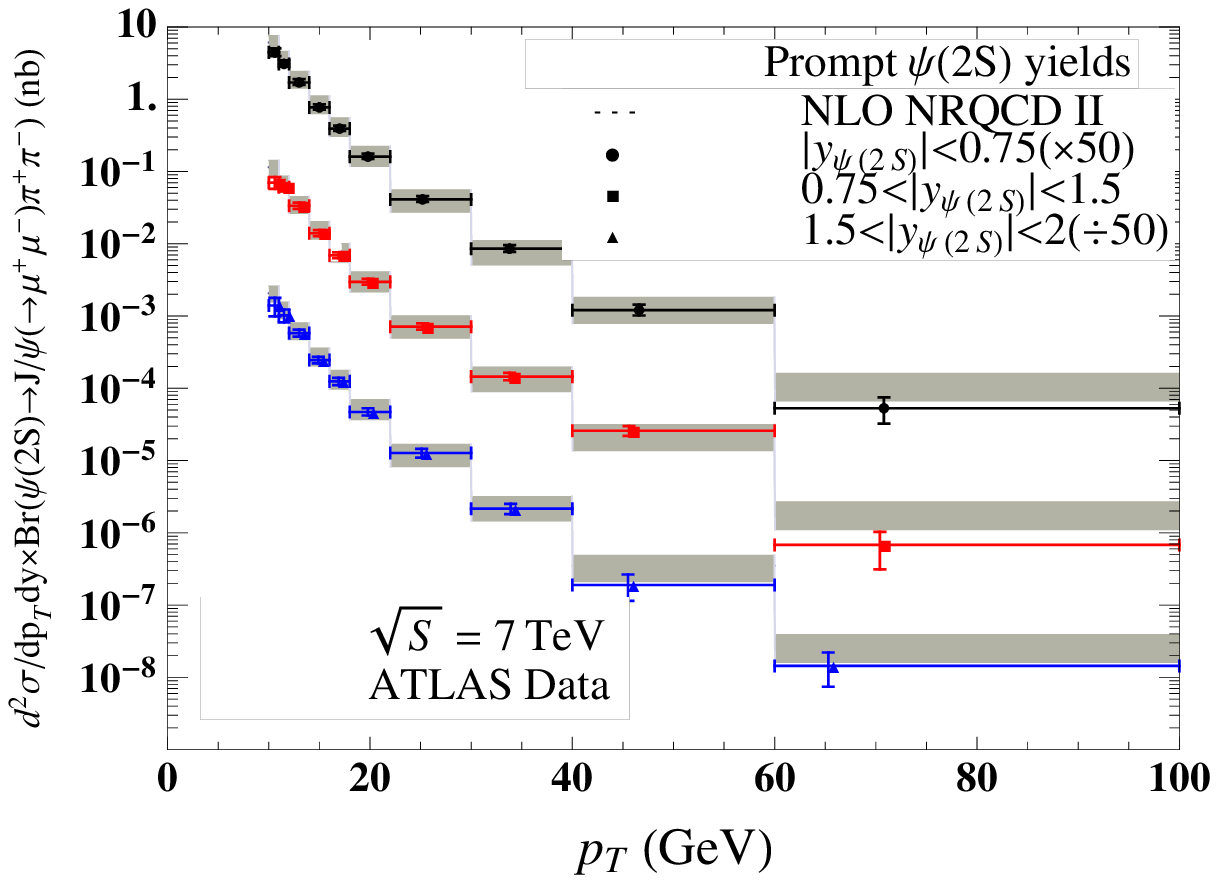}
\caption{\label{fig:psi2syields2}Comparison of NLO NRQCD ( with the set II of CO LDMEs ) and
CDF~\cite{Aaltonen:2009dm},CMS~\cite{Chatrchyan:2011kc},
LHCb~\cite{Aaij:2012ag} and ATLAS~\cite{Aad:2014fpa} data for prompt $\psits$ yields.}
\end{center}
\end{figure}

We update our numerical predictions for $\psits$ yields at the
Tevatron and the LHC as several collaborations have
released their prompt $\psits$ yields
measurements~\cite{Aaltonen:2009dm,Chatrchyan:2011kc,Aaij:2012ag,Aad:2014fpa} in the past a few years. It is
worthwhile to mention that one of the main uncertainty in experimental
measurement comes from the unknown spin-alignment. Hence, it would be
quite useful to give a theoretical prediction on polarisation, which
will be presented in the next subsection. Our NLO NRQCD predictions for
prompt $\psits$ yields are shown in figure \ref{fig:psi2syields} (using default set of CO LDMEs) and figure \ref{fig:psi2syields2} (using set II of CO LDMEs). Our
theoretical results are in good agreement with the experimental data
at the LHC and Tevatron for the regime $p_T>p_{T\rm{cut}}$, where we use $p_{T\rm{cut}}=7\rm{GeV}$ in default set and  $p_{T\rm{cut}}=11\rm{GeV}$ in set II [ Strictly speaking, ATLAS large $p_T$ yields data favor our prediction on set II ]. In the
$p_T<p_{T\rm{cut}}$ regime, experimental data tell us that there
might be a significant non-perturbative smearing effect to violate
the reliability of our fixed-order result.
The error bands in our results represent our theoretical uncertainties, which are
dominated by the uncertainties in CO LDMEs.

\subsection{polarizations}
\begin{figure*}[!hbtp]
\begin{center}
\includegraphics*[scale=0.47]{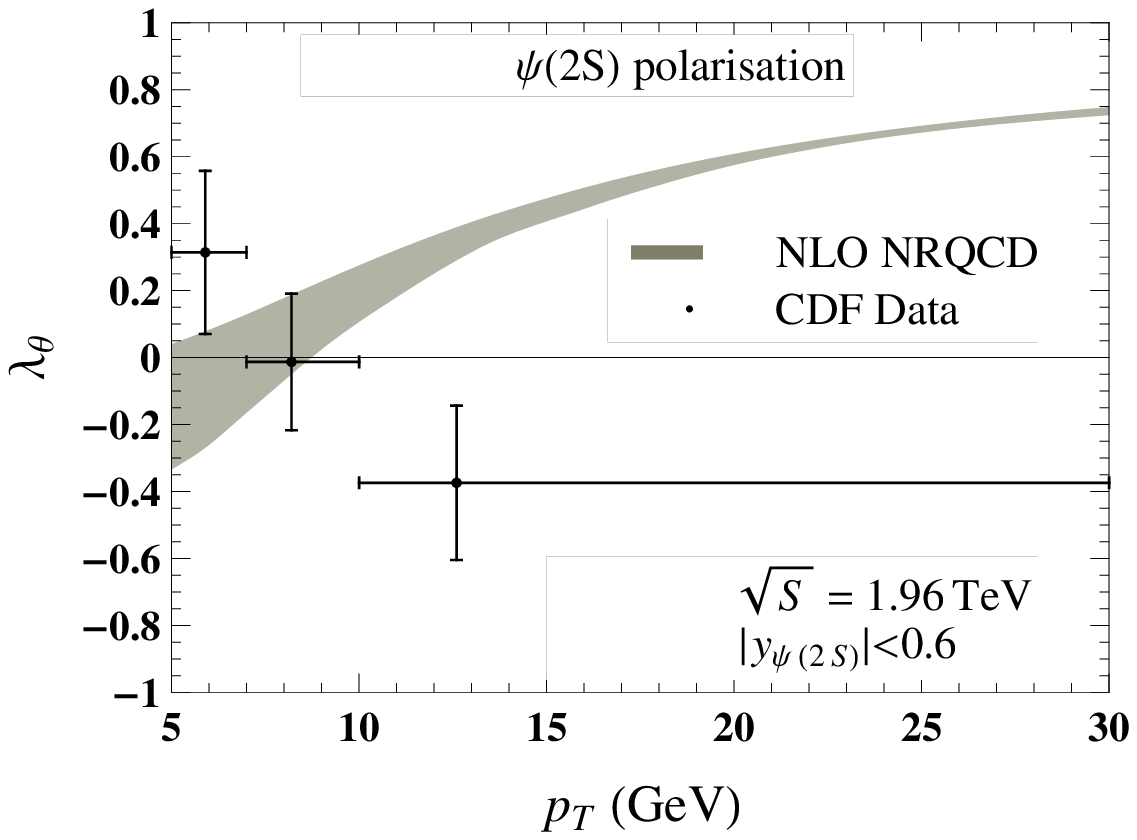}
\includegraphics*[scale=0.47]{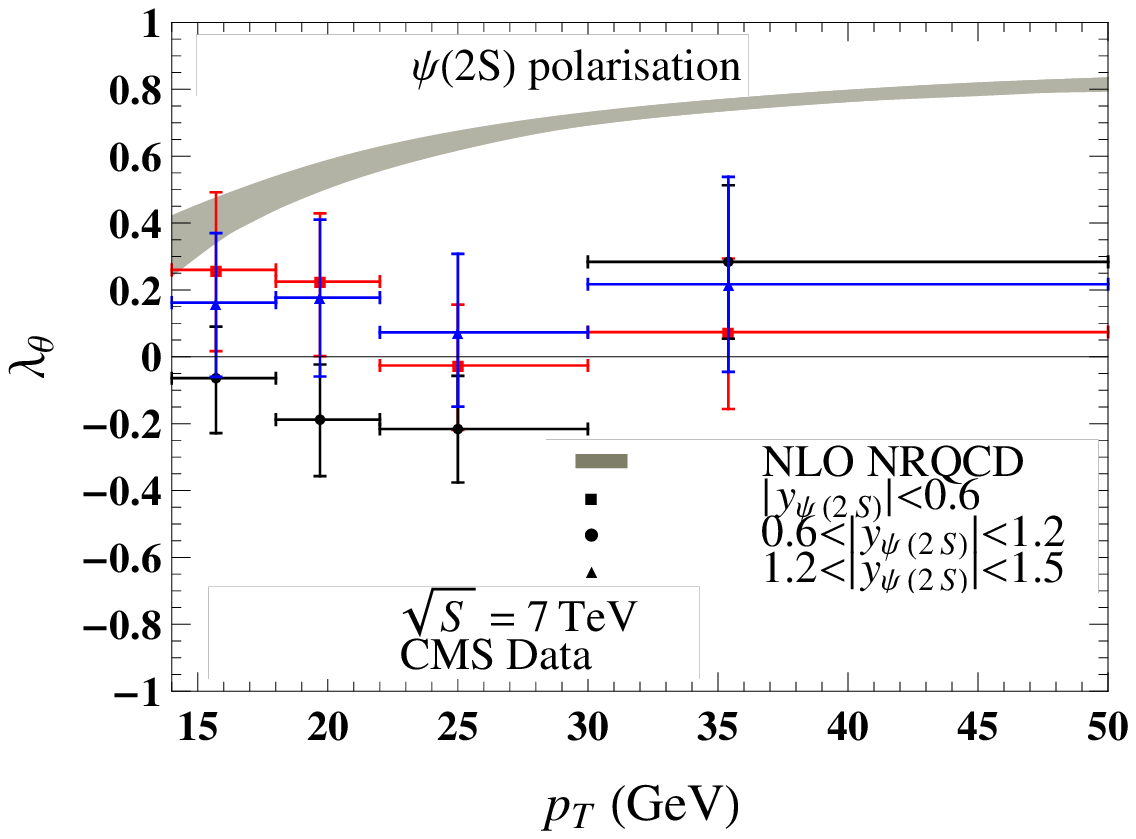}
\includegraphics*[scale=0.47]{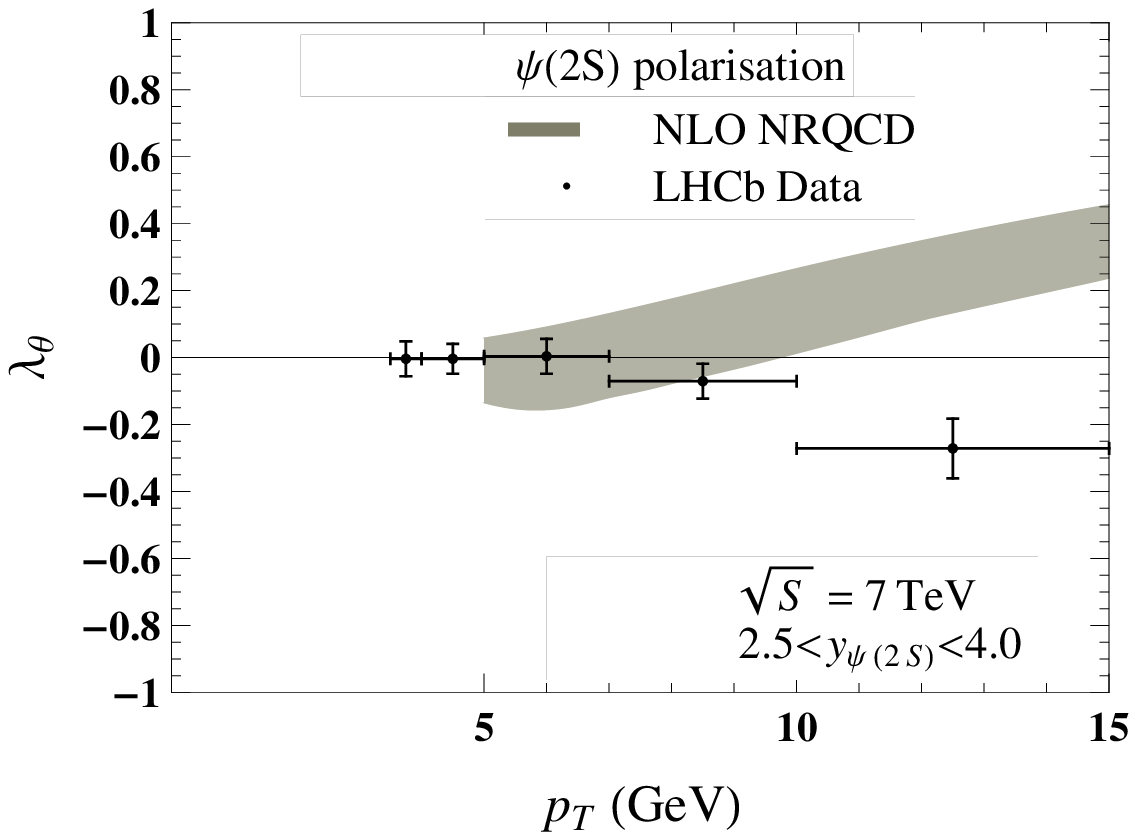}
\caption{\label{fig:psi2spol}Comparison of NLO NRQCD (with the default set of CO LDMEs) and
CDF~\cite{Abulencia:2007us}, CMS~\cite{Chatrchyan:2013cla} and LHCb~\cite{Aaij:2014qea} data
for prompt $\psits$ polarisation $\lambda_{\th}$ in helicity frame.}
\end{center}
\end{figure*}

\begin{figure*}[!hbtp]
\begin{center}
\includegraphics*[scale=0.47]{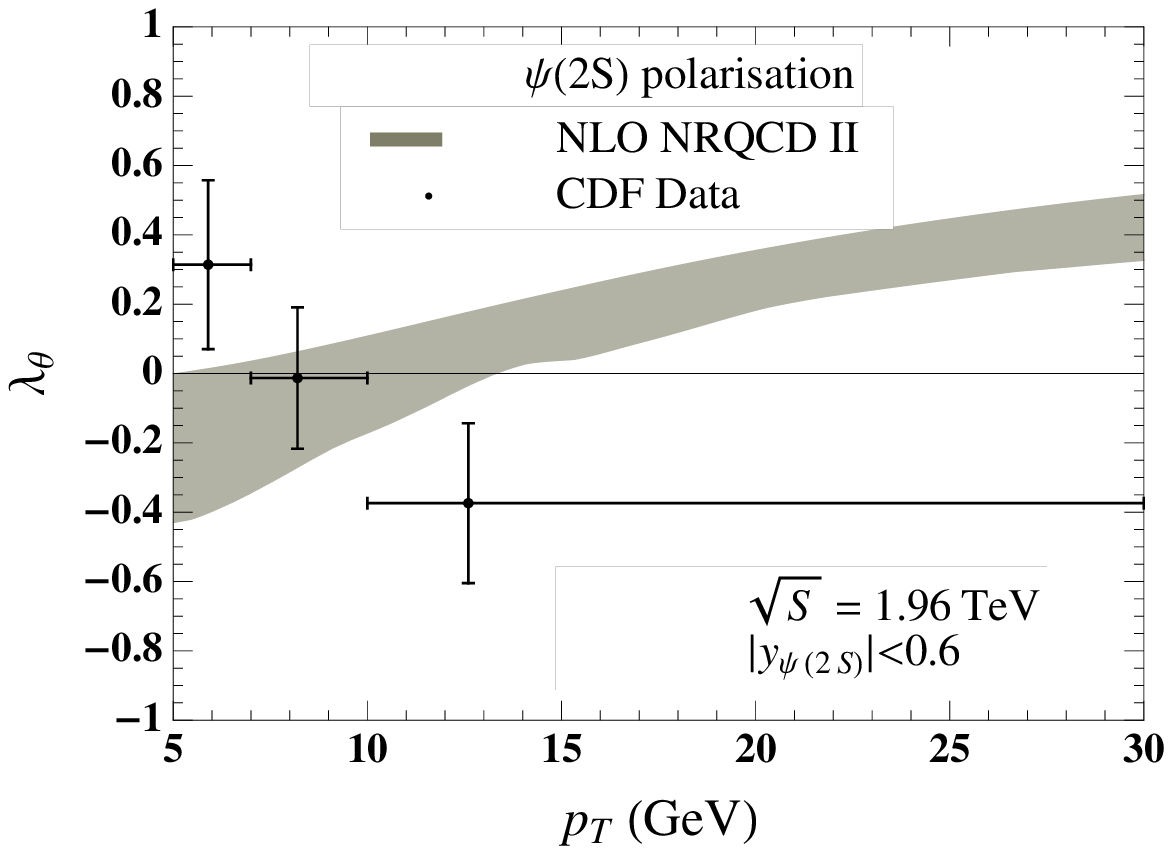}
\includegraphics*[scale=0.47]{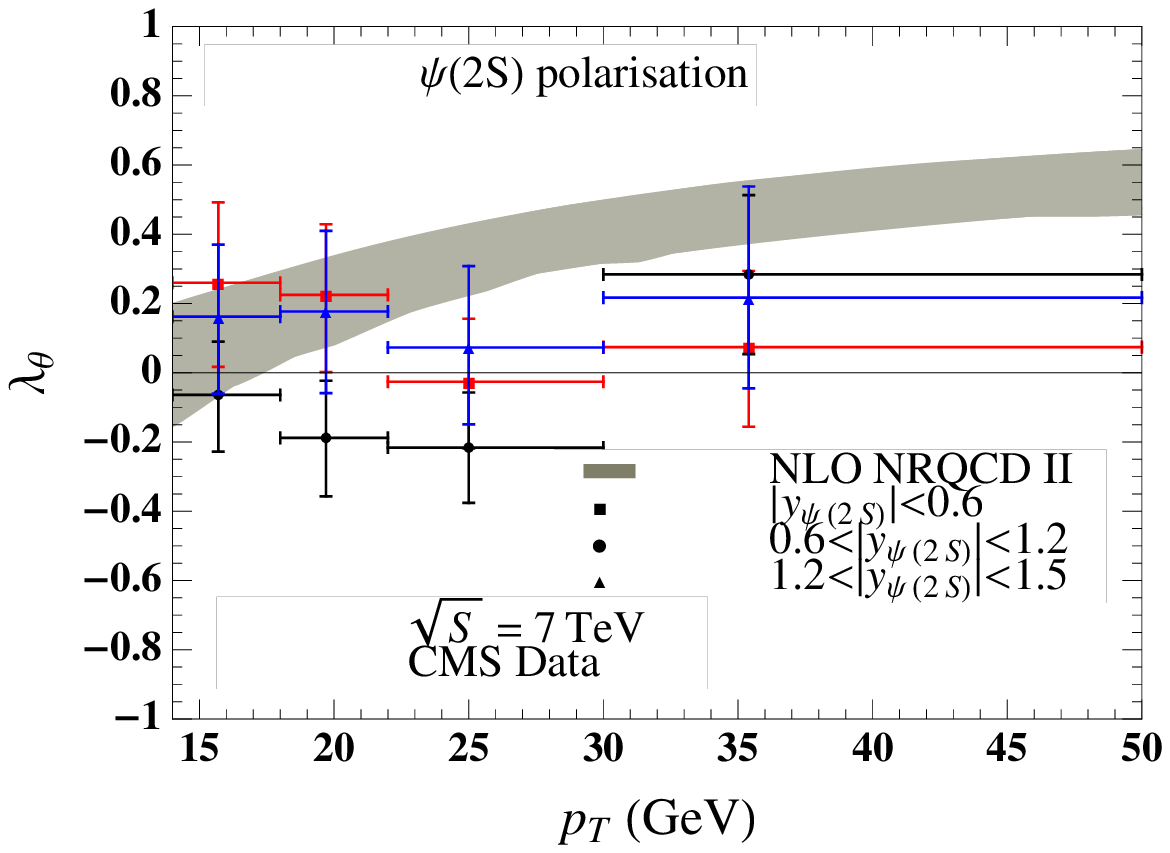}
\includegraphics*[scale=0.47]{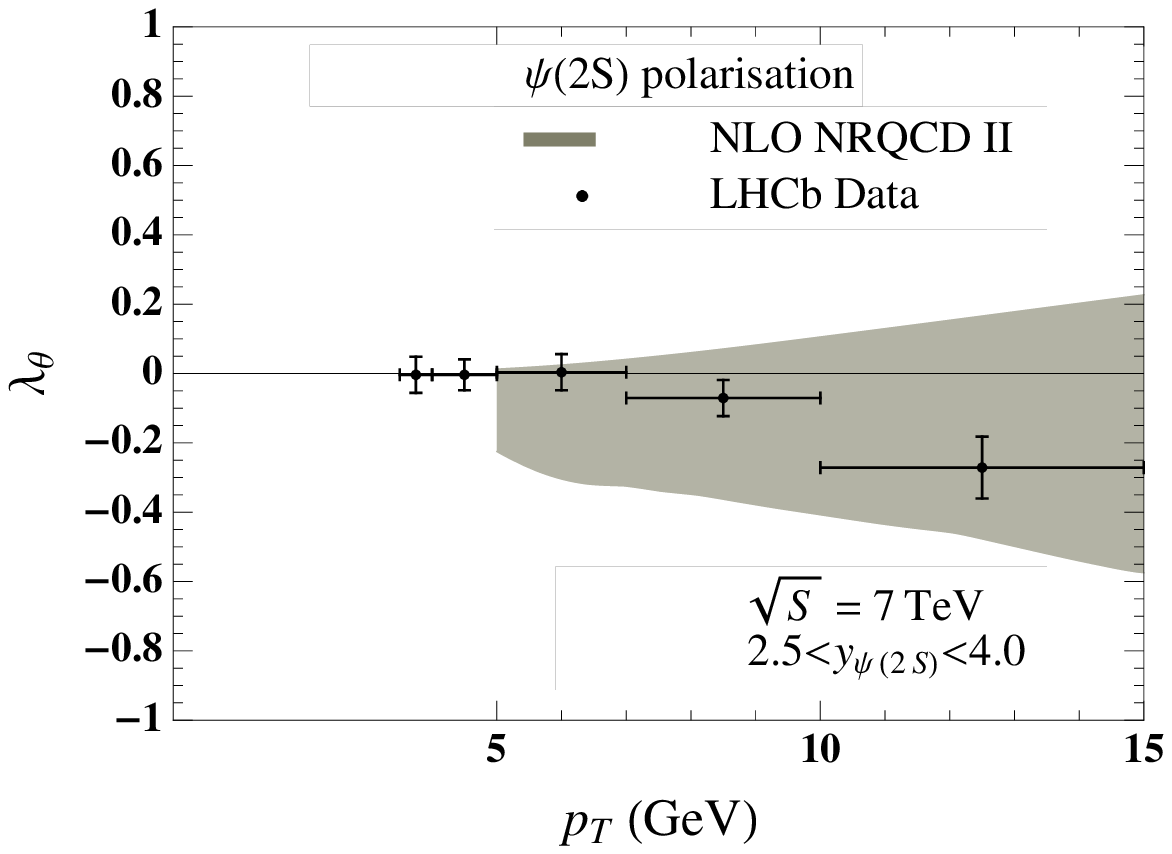}
\caption{\label{fig:psi2spol2}Comparison of NLO NRQCD (with the set II of CO LDMEs) and
CDF~\cite{Abulencia:2007us}, CMS~\cite{Chatrchyan:2013cla} and LHCb~\cite{Aaij:2014qea} data
for prompt $\psits$ polarisation $\lambda_{\th}$ in helicity frame.}
\end{center}
\end{figure*}

We are in the position to give the theoretical predictions of
polarisation observable $\lambda_{\th}$ for prompt $\psits$. We
compare our NLO NRQCD results with the experimental data given by
CDF~\cite{Abulencia:2007us} and CMS~\cite{Chatrchyan:2013cla}
collaborations in figure \ref{fig:psi2spol} (using default set of CO LDMEs) and figure \ref{fig:psi2spol2} (using set II of CO LDMEs). As we discussed in section \ref{sec:ldmes}, a larger value of $\Mbpsits$
will result in a larger transverse component for prompt $\psits$. Hence, using our default set of CO LDMEs, the resulted $\lambda_{\th}$ are much larger than the data (see figure \ref{fig:psi2spol}), while values of $\lambda_{\th}$ calculated by using set II of CO LDMEs in figure \ref{fig:psi2spol2} can marginally describe the data. On the experimental side, there seems to be a little bit inconsistence between the CDF~\cite{Abulencia:2007us}
data and the CMS~\cite{Chatrchyan:2013cla} data, and the error bars
are large. Therefore, a more precise measurement at the LHC is essential to
clarify the difference in the future.

We would emphasize here that if we simply set  $\Mbpsits$ to be zero and only keep $\sps$, it will of course
result in unpolarized results for any polarisation observables in
any frame, which is also noticed in refs.~\cite{Ma:2010yw,Faccioli:2014cqa}.

\section{Prompt $\jpsi$ yields and polarizations\label{sec:4}}
The prompt $\jpsi$ production in hadronic collisions is more
involved. It receives a significant contribution from $\chi_{cJ}$
and $\psits$ decay via $\chi_{cJ}\to\jpsi+\gamma$ and
$\psits\to\jpsi+X$ respectively, which is usually called the
feeddown contribution. $\jpsi$ can be reconstructed quite well from
its decay products, a muon pair. In our previous
study~\cite{Chao:2012iv}, we did not include feeddown contribution
in our $\jpsi$ yields and polarisation predictions. We found there
was still a parameter space for CO LDMEs to give an almost
unpolarized theoretical prediction, though we were still unable to
extract the three independent CO LDMEs unambiguously. More precisely, we need a cancellation happens between the transverse
components of $\sb$ and $\pj$ to give an unpolarized result, which happens to be equivalent to need a (very) small $\Mbjpsi$. Later, we also
consider the impact of feeddown contribution from $\chi_{cJ}$ decay
on our direct $\jpsi$ polarisation~\cite{Shao:2012fs,Shao:2014fca}. From eq.~(C4)
in ref.~\cite{Shao:2012fs}, the feeddown contribution from
$\chi_{c1}$ for $\jpsi$ polarisation is in the interval
$[-\frac{1}{3},1]$, while the feeddown contribution from $\chi_{c2}$
is in the interval $[-\frac{3}{5},1]$, regardless of its production
mechanism.\footnote{The $\jpsi$ polarisation $\lambda_{\th}$ from
scalar particle $\chi_{c0}$ is always zero.} We showed that the
smearing from feeddown contribution will not change our result too
much based on our direct $\jpsi$ polarisation. Now, we are intending
to give a rigorous prediction for prompt $\jpsi$ yields and
polarisation after including the feeddown contribution from
$\chi_{cJ}$ and $\psits$ decay. As we discussed in section \ref{sec:ldmes}, the LDMEs of $\Majpsi$ and $\Mbjpsi$ are insensitive to the $p_{T\rm{cut}}$ when $p_{T\rm{cut}}>7\rm{GeV}$. We will use the values of $\Majpsi$ and $\Mbjpsi$ obtained from $p_T>7\rm{GeV}$ data only in this section.

\subsection{yields}

\begin{center}
\begin{figure}
\hspace{0cm}\includegraphics[width=0.45\textwidth]{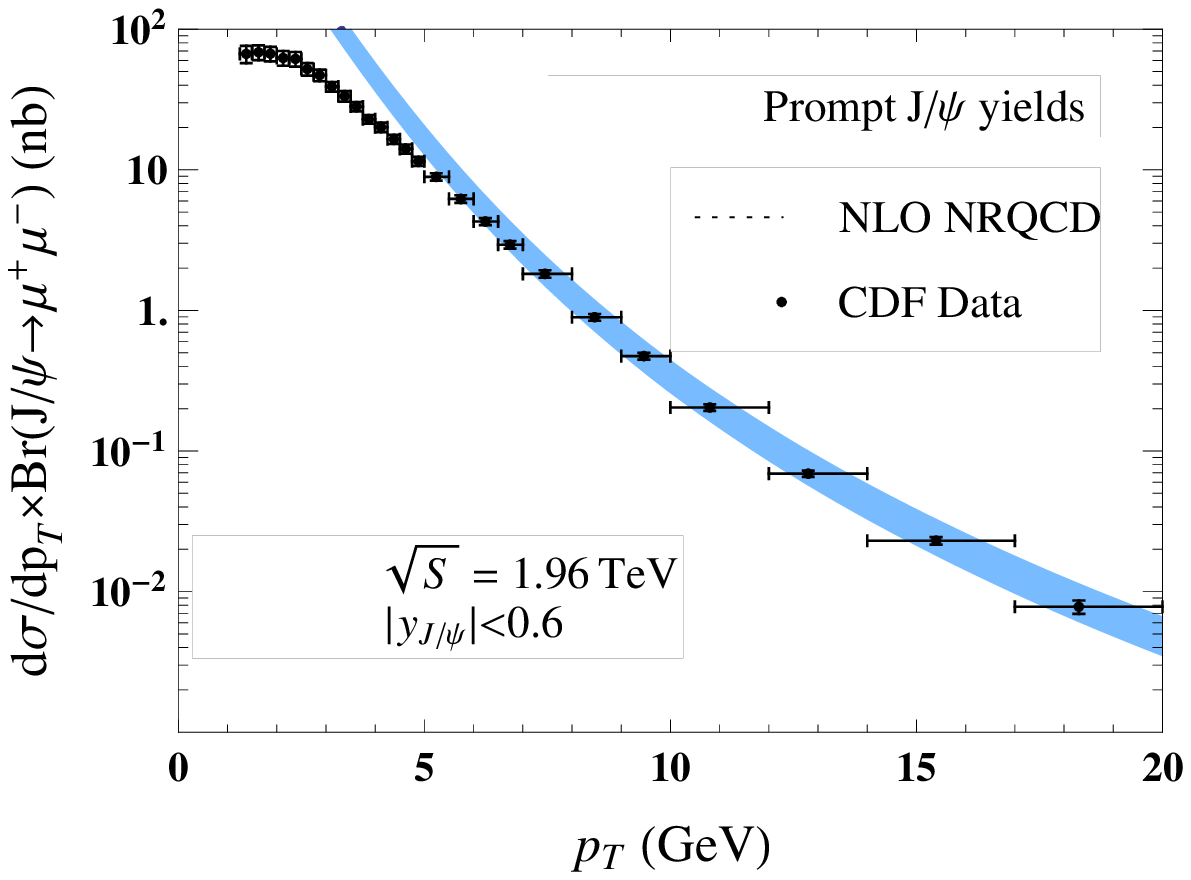}
\hspace{0cm}\includegraphics[width=0.45\textwidth]{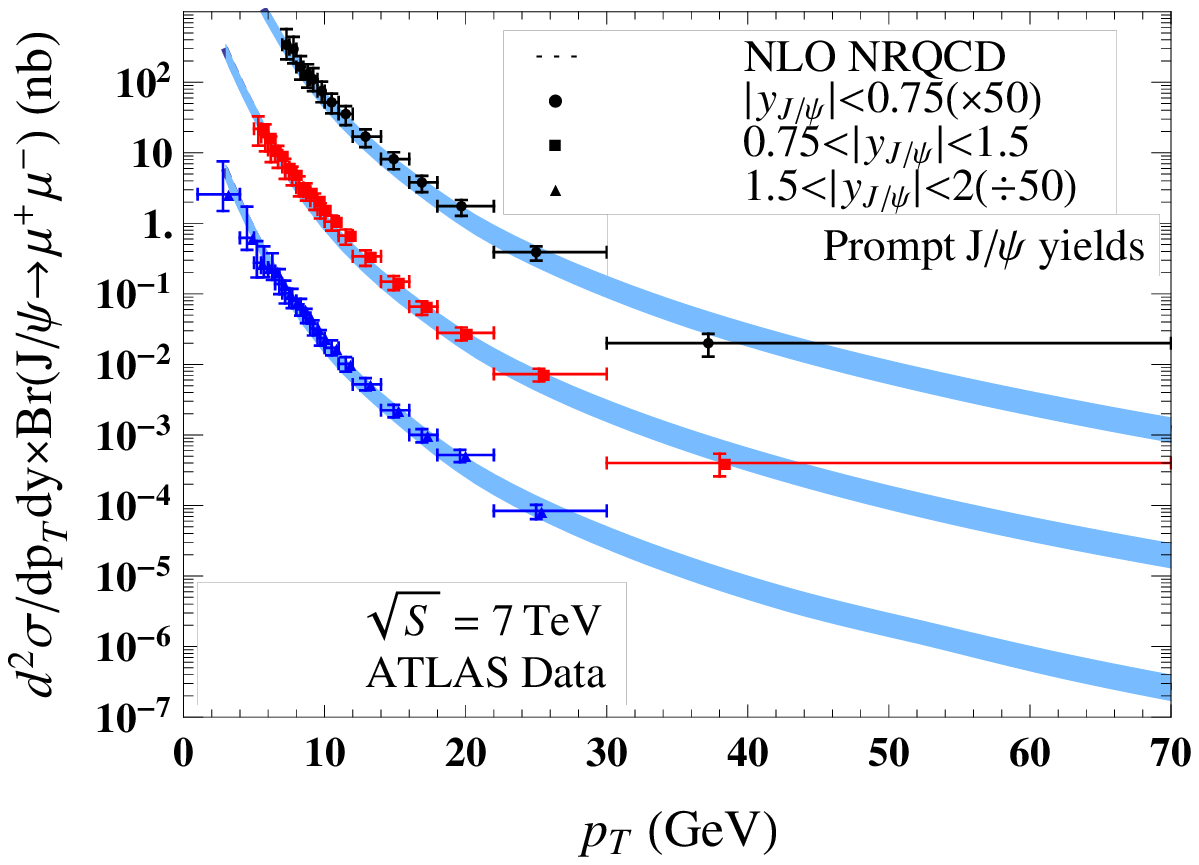}
\hspace{1cm}\includegraphics[width=0.45\textwidth]{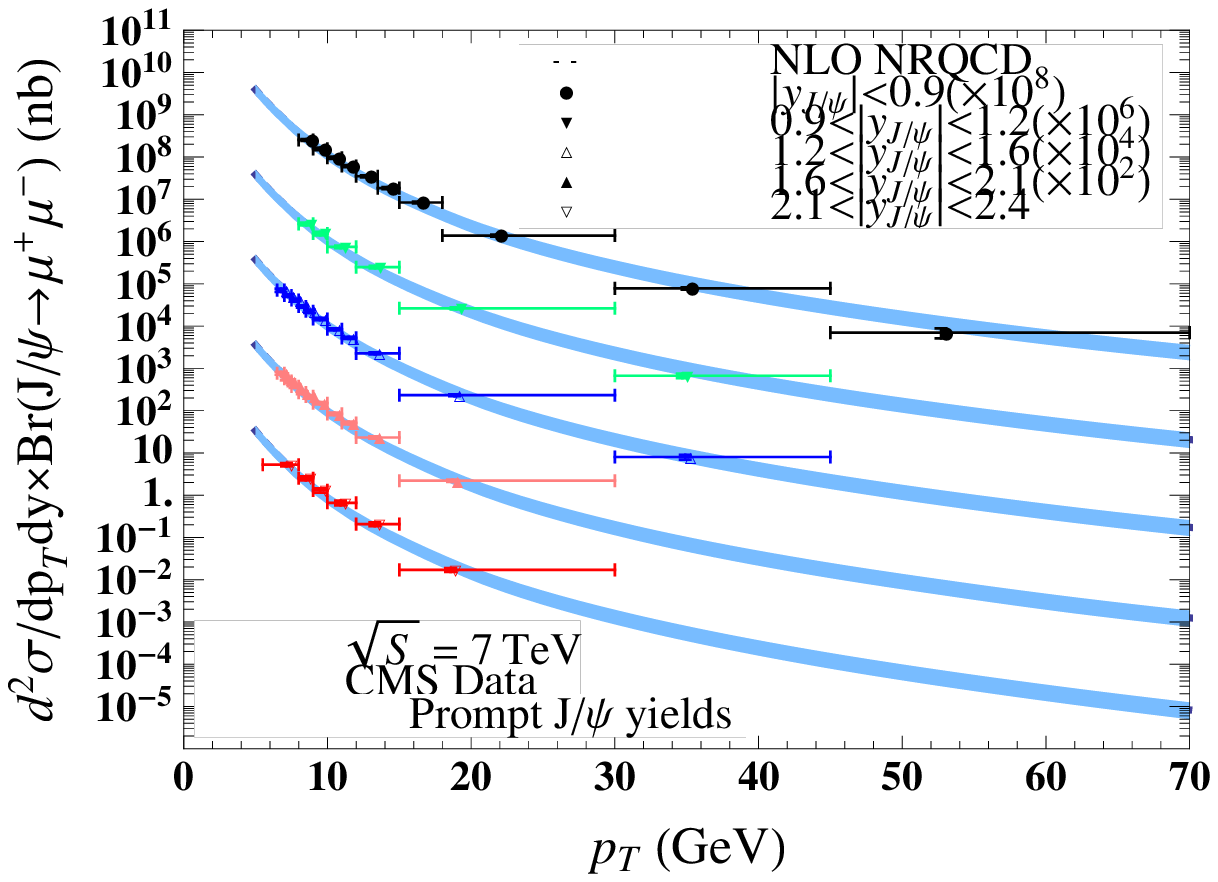}
\hspace{1.4cm}\includegraphics[width=0.45\textwidth]{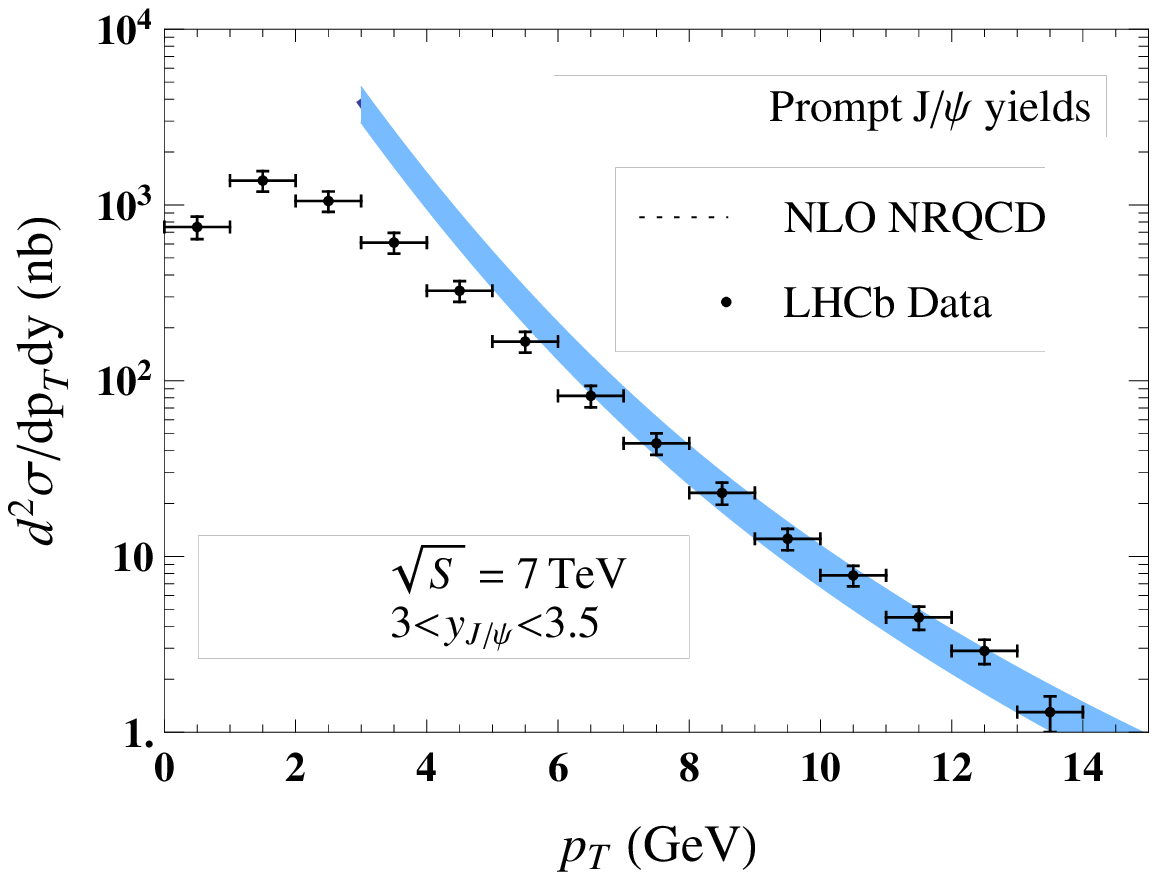}
\caption{\label{fig:jpsiyields}Comparison of NLO NRQCD and
CDF~\cite{Acosta:2004yw},ATLAS~\cite{Aad:2011sp},CMS~\cite{Chatrchyan:2011kc}
and LHCb~\cite{Aaij:2011jh} data for prompt $\jpsi$ yields.}
\end{figure}
\end{center}

In this subsection, we present the $p_T$ spectrum for prompt $\jpsi$
yields. We show our NLO NRQCD predictions for prompt $\jpsi$ yields
in figure~\ref{fig:jpsiyields}. The experimental data are taken from
CDF~\cite{Acosta:2004yw}, ATLAS~\cite{Aad:2011sp}, CMS~\cite{Chatrchyan:2011kc}
and LHCb~\cite{Aaij:2011jh}. Good agreement is found up to 70 GeV
and in various rapidity bins.

\begin{center}
\begin{figure}
\hspace{0cm}\includegraphics[width=0.45\textwidth]{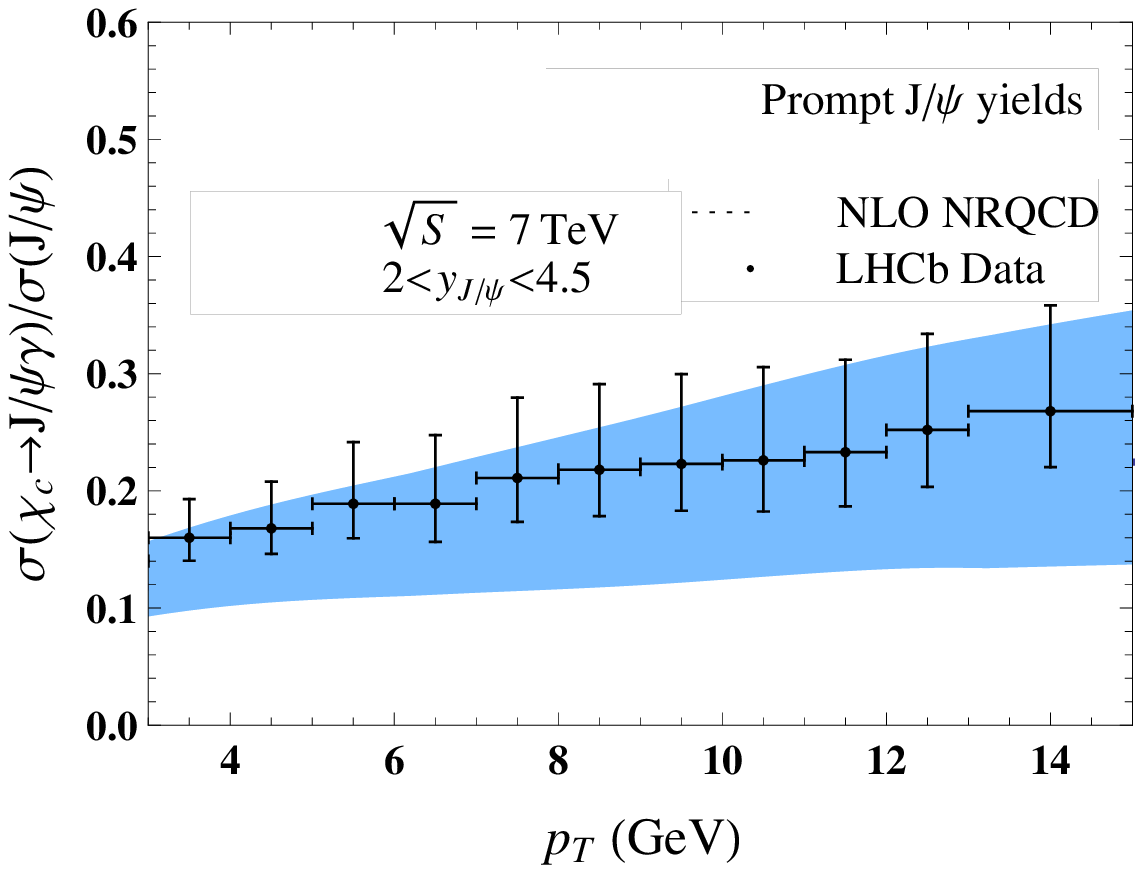}
\hspace{1cm}\includegraphics[width=0.45\textwidth]{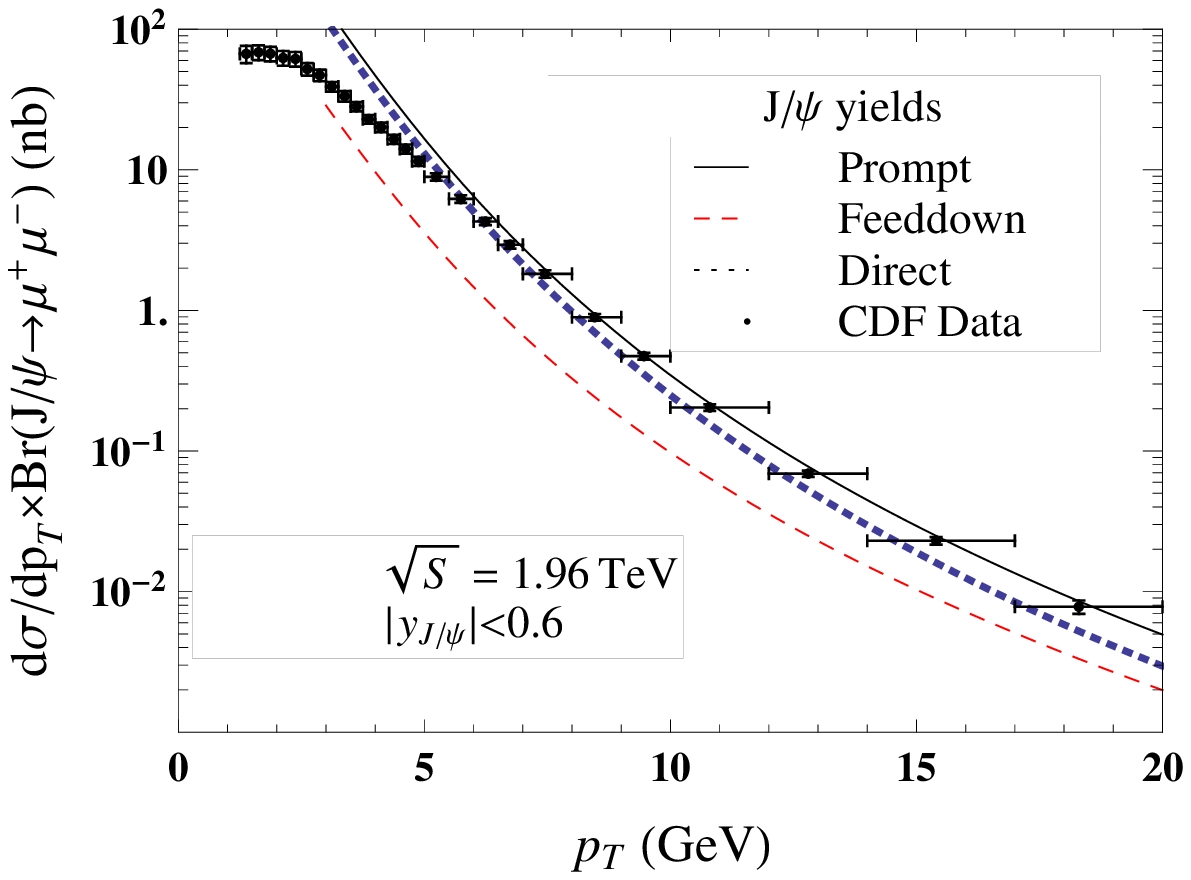}
\caption{\label{fig:fdjpsiyields}Comparison of NLO NRQCD and
LHCb~\cite{LHCb:2012af} and
CDF~\cite{Acosta:2004yw} data for $\jpsi$ yields.}
\end{figure}
\end{center}

\begin{center}
\begin{figure}
\hspace{0cm}\includegraphics[width=0.45\textwidth]{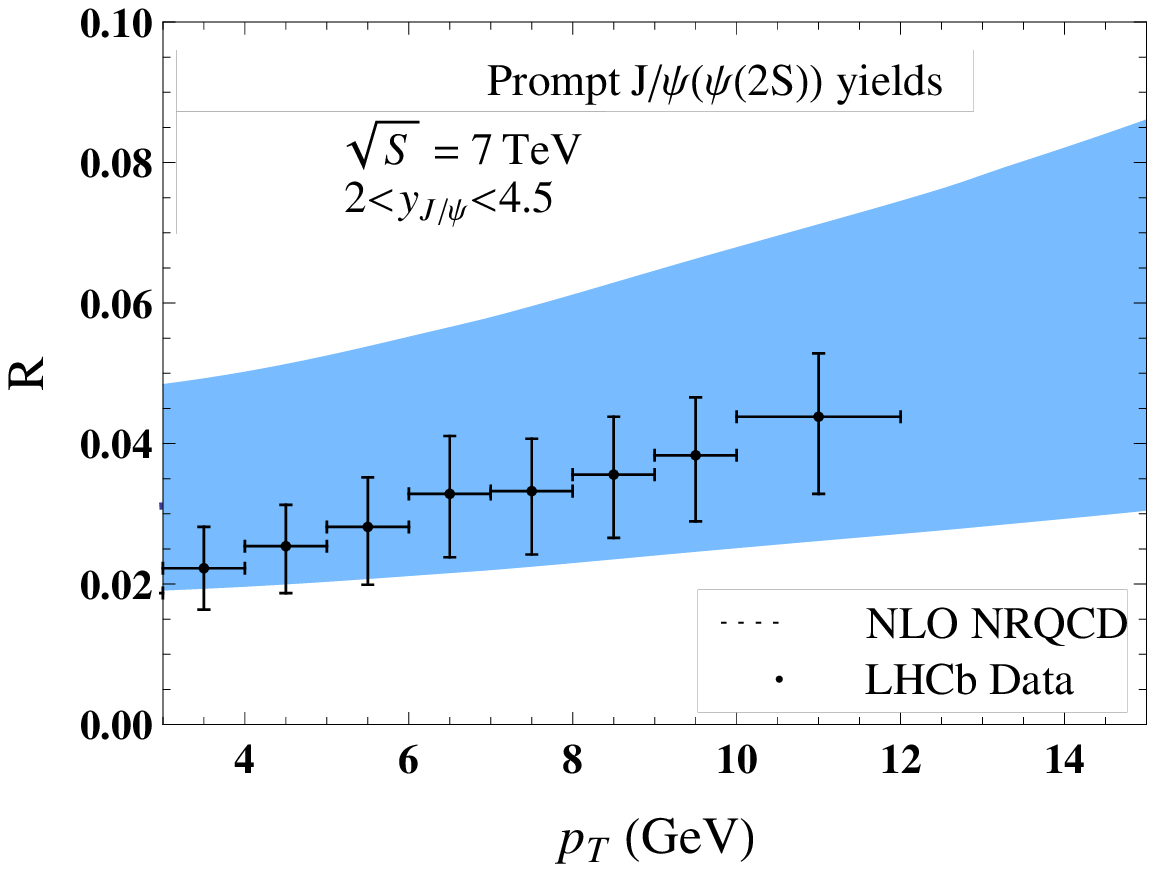}
\hspace{1cm}\includegraphics[width=0.45\textwidth]{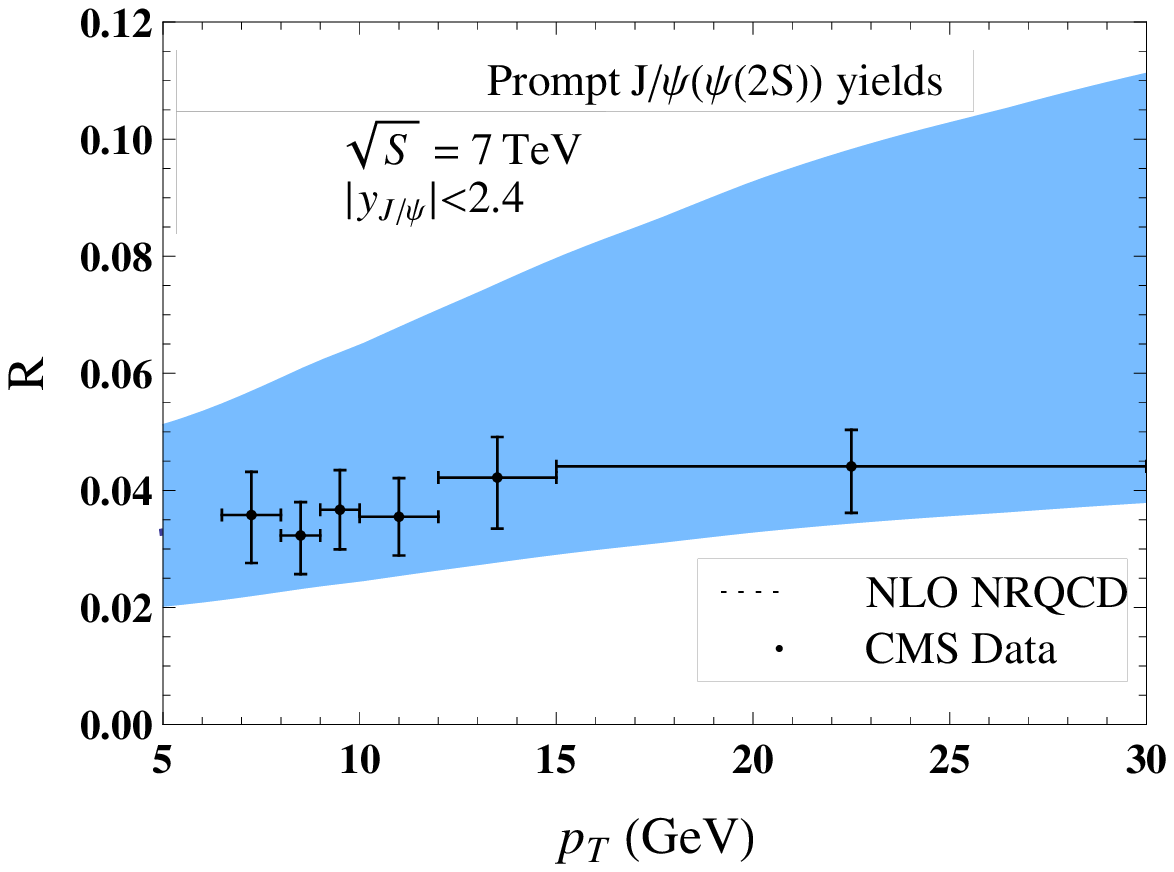}\\
\hspace{0cm}\includegraphics[width=0.45\textwidth]{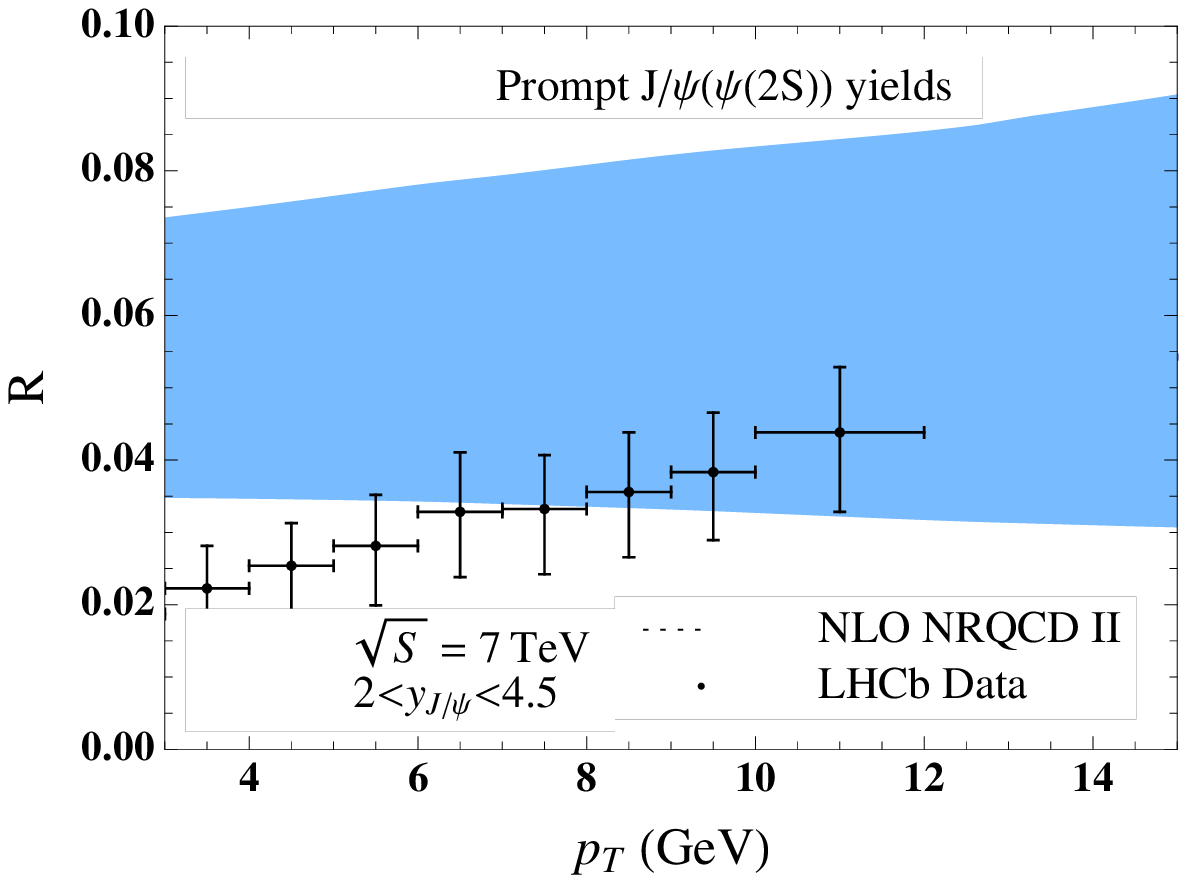}
\hspace{1cm}\includegraphics[width=0.45\textwidth]{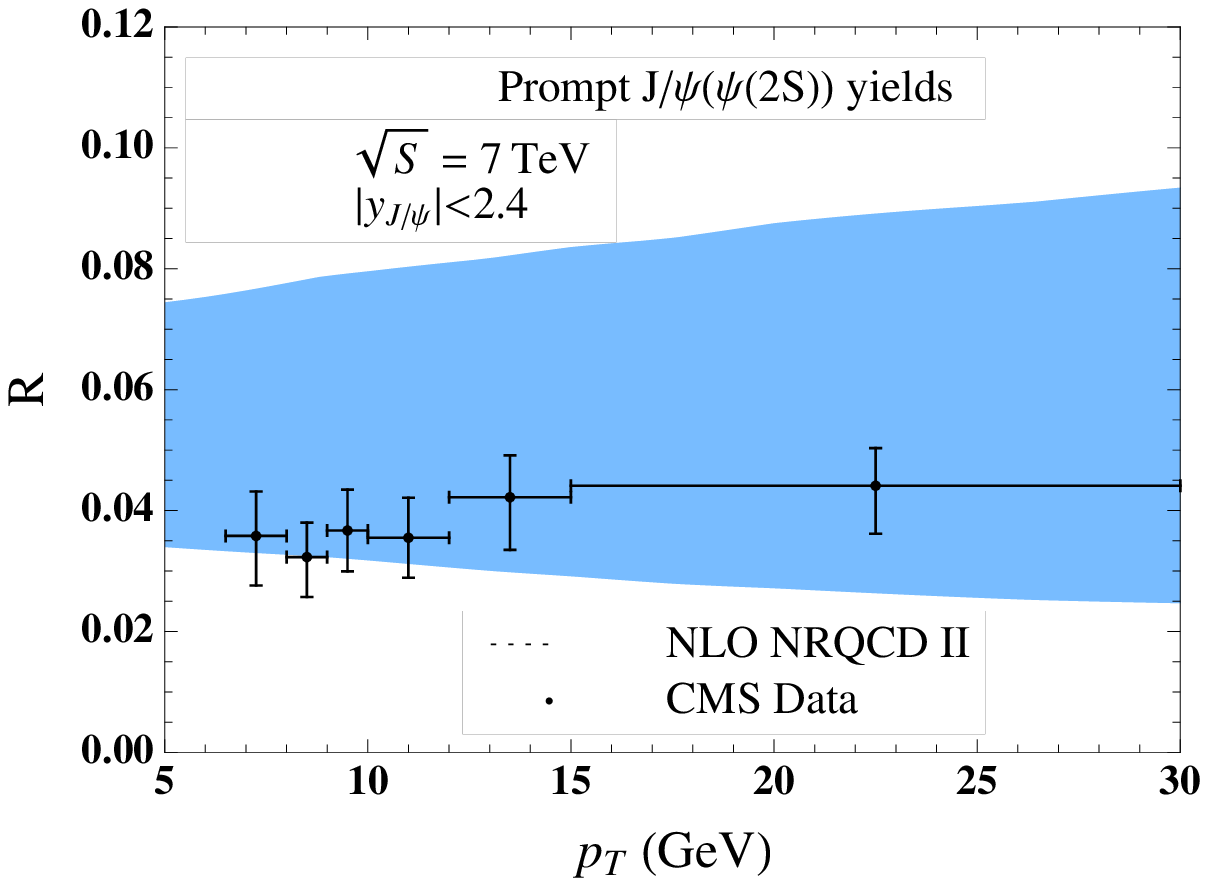}
\caption{\label{fig:Rjpsiyields}Comparison of NLO NRQCD and
LHCb~\cite{Aaij:2012ag} and CMS~\cite{Chatrchyan:2011kc} data for $R$. We use the default set of CO LDMEs for $\psits$ in the upper two panels, while the lower two panels are obtained by using the set II of CO LDMEs for $\psits$.}
\end{figure}
\end{center}

In order to understand the fraction of feeddown contribution from
$\chi_{cJ}$ to prompt $\jpsi$, we also show the theoretical
prediction for $\frac{\sigma(\chi_c\to\jpsi\gamma)}{\sigma(\jpsi)}$
in figure~\ref{fig:fdjpsiyields} in the LHCb fiducial region. The plot
implies that the $p_T$ spectrum of prompt $\chi_c$ is harder than
that of $\jpsi$, which can be understood as $\chi_c$ has a stronger $p_T^{-4}$ behaviour. In figure \ref{fig:Rjpsiyields}, we also show the ratio $R$ of prompt $\psits$ yields and
prompt $\jpsi$ yields as defined in
refs.~\cite{Aaij:2012ag,Chatrchyan:2011kc}, \bqa
R\equiv\frac{\sigma(\psits\to\mu^+\mu^-)}{\sigma(\jpsi\to\mu^+\mu^-)},
\eqa which indicates the $p_T$ dependence of feeddown contribution
from $\psits$ in prompt $\jpsi$ yields. With the default set of CO LDMEs for $\psits$, it increases as $p_T$
becomes larger because of $\Mbpsits/\Mapsits > \Mbjpsi/\Majpsi$. On the contrast, after using the new set II of CO LMDEs for $\psits$, the ratio $R$ is flat in $p_T$, which is easily understood because of a smaller $\Mbpsits/\Mapsits$. Finally, we divide the prompt $\jpsi$ yields into direct $\jpsi$
yields and the feeddown $\jpsi$ from $\chi_c$ and $\psits$ decay in
the second plot of figure~\ref{fig:fdjpsiyields}. It shows that
the $p_T$ spectrum of feeddown $\jpsi$ is harder than that of direct
one.

Before going ahead into the discussion of the polarization case, we want to clarify that only the ratio $R$ is sensitive to different sets of CO LMDEs for $\psits$ in this subsection, while other differential distributions are not. It is just because the feed-down contribution from  $\psits$ in prompt $\jpsi$ production is indeed small. This fact has also been checked numerically.  It is also applicable to the polarization observable $\lambda_{\th}$ for prompt $\jpsi$ in the next subsection. Hence, we will refrain ourselves from presenting the similar plots by using the set II of CO LDMEs for $\psits$ except the ratio $R$.

\subsection{polarizations}
\begin{center}
\begin{figure}
\hspace{0cm}\includegraphics[width=0.45\textwidth]{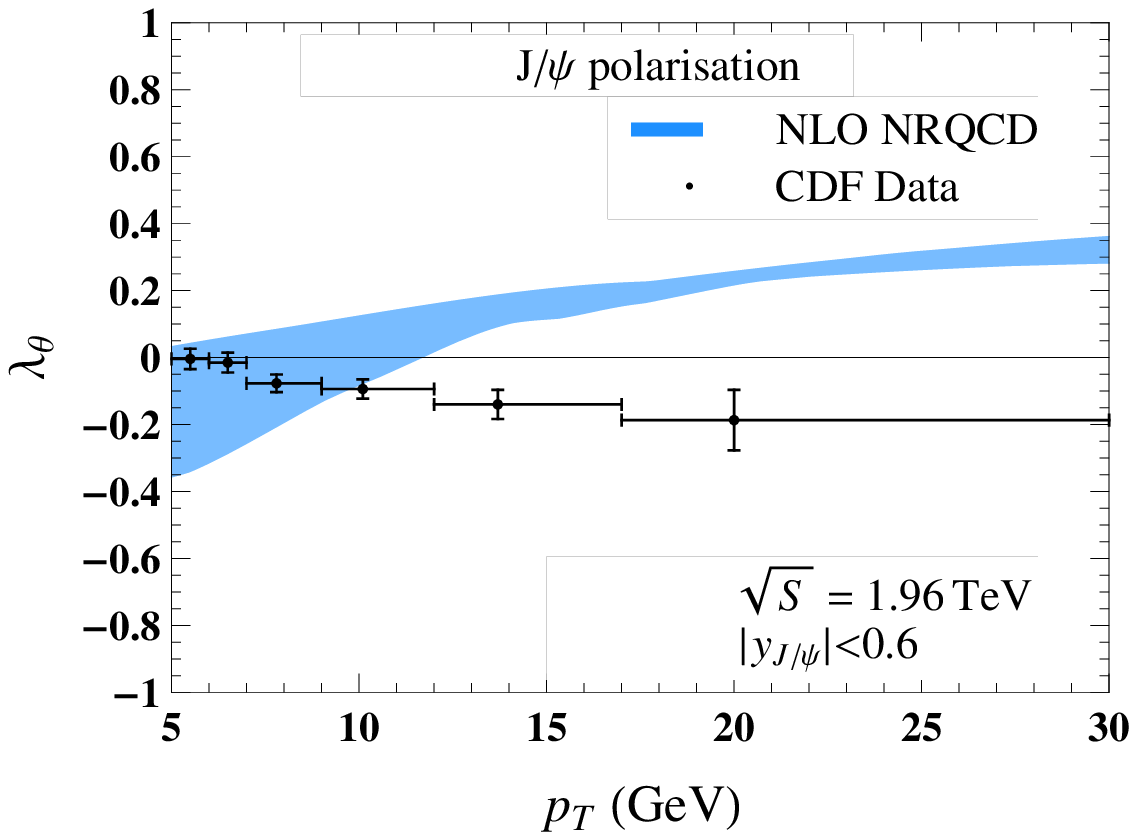}
\hspace{1cm}\includegraphics[width=0.45\textwidth]{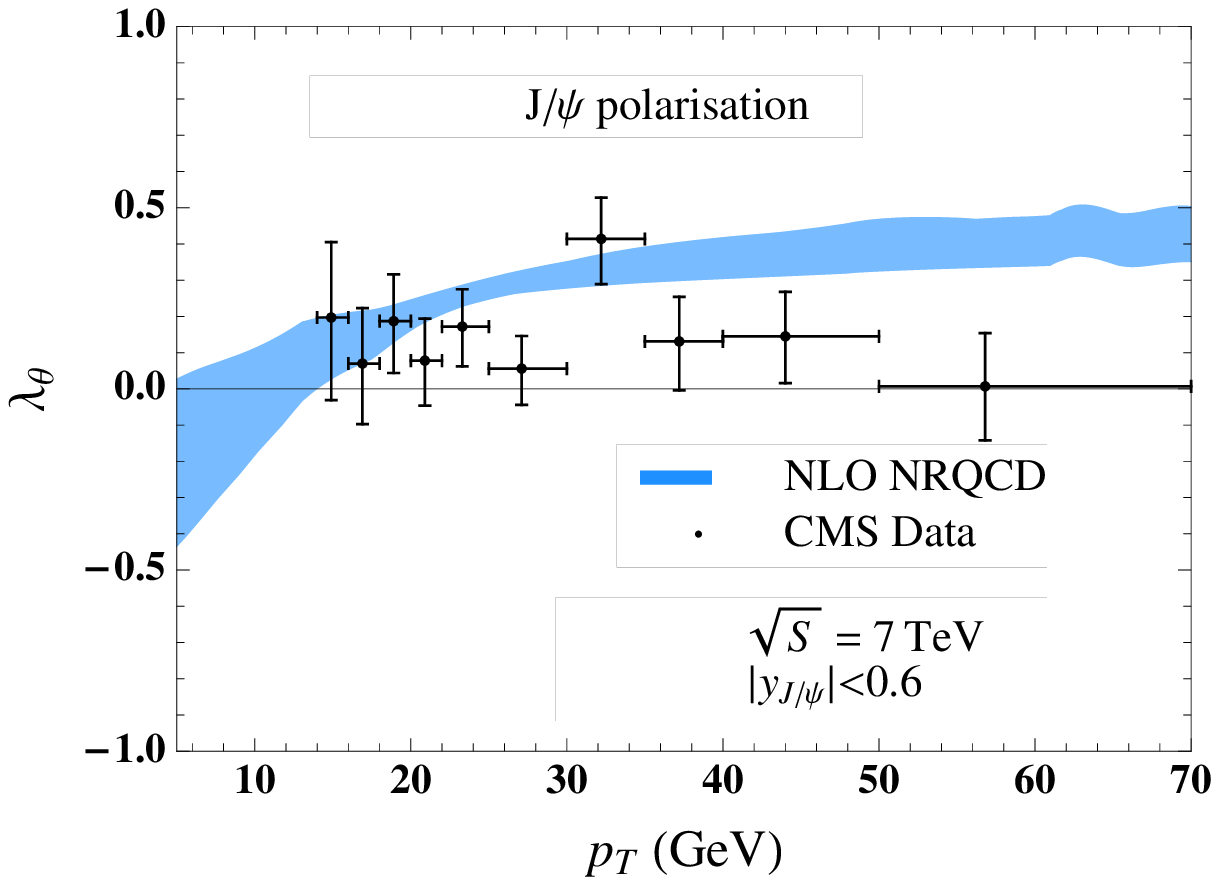}
\hspace{1cm}\includegraphics[width=0.45\textwidth]{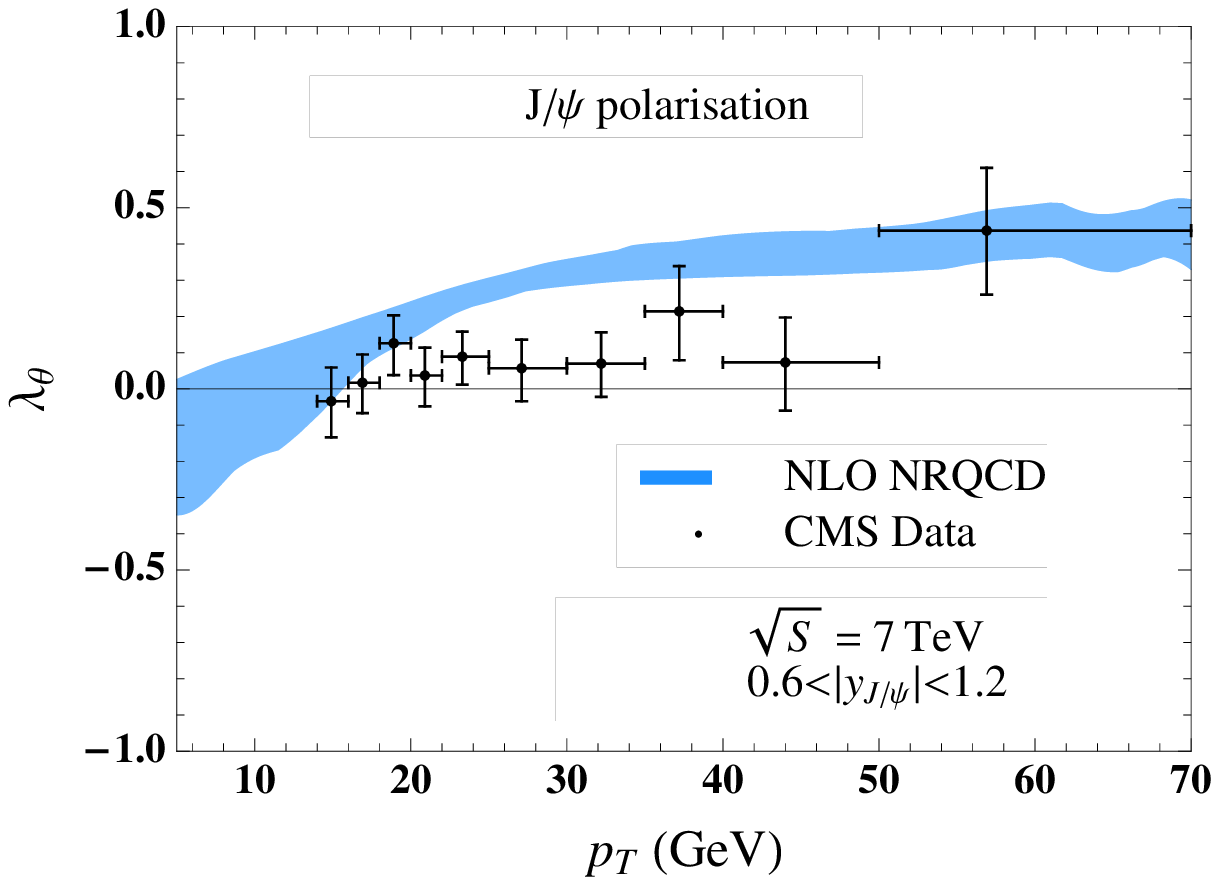}
\hspace{1.4cm}\includegraphics[width=0.45\textwidth]{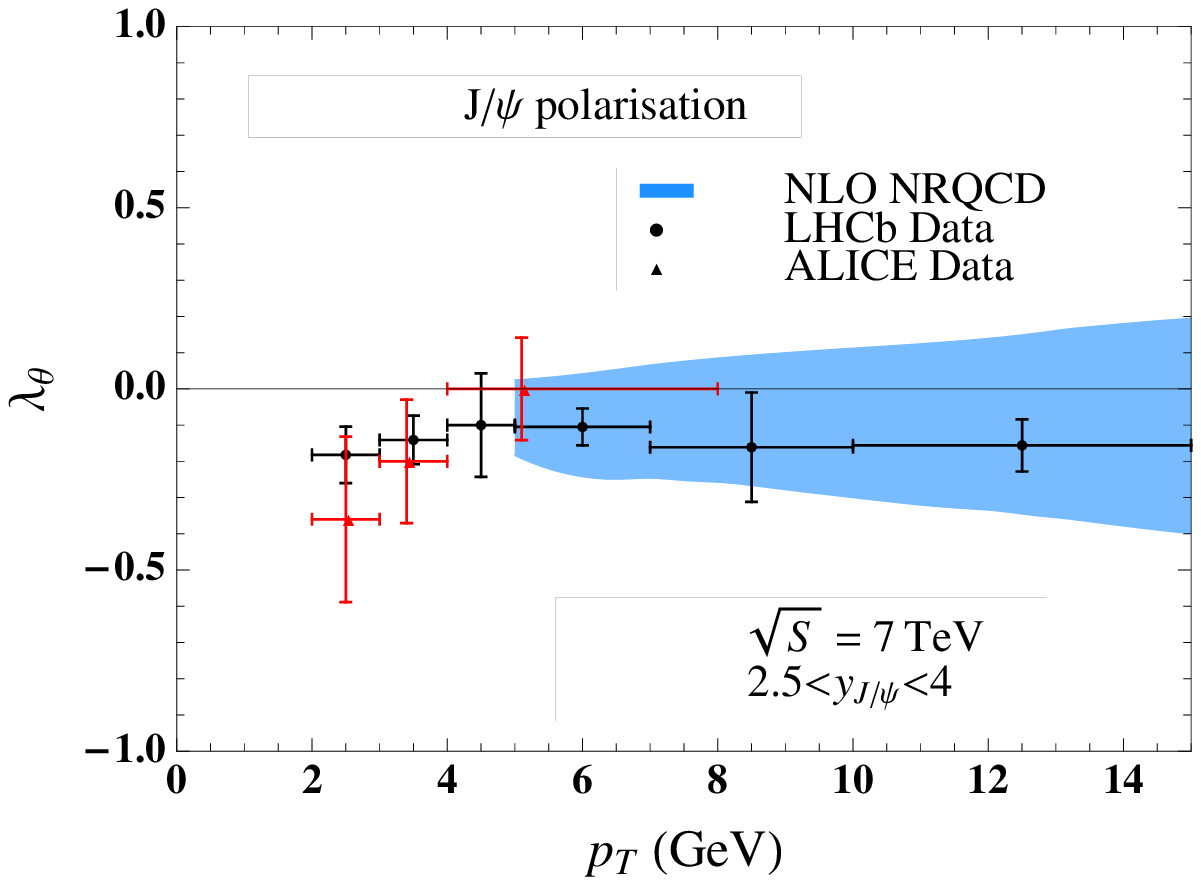}
\caption{\label{fig:jpsipol}Comparison of NLO NRQCD and
CDF~\cite{Abulencia:2007us},CMS~\cite{Chatrchyan:2013cla},
LHCb~\cite{Aaij:2013nlm} and ALICE~\cite{Abelev:2011md} data for
prompt $\jpsi$ polarisation $\lambda_{\th}$ in helicity frame. The
ALICE~\cite{Abelev:2011md} data is for the inclusive $\jpsi$.}
\end{figure}
\end{center}

The polarisation for prompt $\jpsi$ should be expected to be almost unpolarized
because a smaller $\Mbjpsi$ indicates
a smaller transverse polarized component in prompt $\jpsi$. We
compare our NLO NRQCD results with
CDF~\cite{Abulencia:2007us}, CMS~\cite{Chatrchyan:2013cla},
LHCb~\cite{Aaij:2013nlm} and ALICE~\cite{Abelev:2011md} data in
figure~\ref{fig:jpsipol}. $\lambda_{\th}$ in different rapidity bins
are close to 0, which is consistent with our previous claim even
after including feeddown
contribution~\cite{Chao:2012iv,Shao:2012fs}. Our results are in good
agreement with the measurements of
CMS~\cite{Chatrchyan:2013cla},\footnote{Although there seems to be some difference between our theoretical results and the current CMS polarization data, we would like to mention that there are still some statistical fluctuations in the CMS data themselves, such as shown in the last bins in $|y|<0.6$ and $0.6<|y|<1.2$ in figure \ref{fig:jpsipol}.} LHCb~\cite{Aaij:2013nlm} and
ALICE~\cite{Abelev:2011md} collaborations, while it is not so good
with CDF data~\cite{Abulencia:2007us}. However, it is worthwhile to
note that the CDF data is also inconsistent with the CMS data in
the same rapidity interval.

Our positive LDMEs assumption is consistent with experiment that
the LHCb data is a little bit lower than the CMS data. As we have pointed out in section~\ref{sec:2}, positivity of LDMEs implies that the $\lambda_{\th}$ will be smaller in the forward
rapidity bin than in the central rapidity bin, based on the
understanding that $\Mbjpsi$ is smaller when rapidity $y$ is larger.
On the other hand, there are negative values of $\mopc$ in the other
two groups~\cite{Butenschoen:2012px,Gong:2012ug}. They will give
larger values of $\lambda_{\th}$ in the forward rapidity bins, which
will be in conflict with the LHCb data.

\section{Summary\label{sec:5}}

With large samples of heavy quarkonium accumulated at the LHC,
quarkonium physics has reached the precision era even at large
transverse momenta regime. In this article, we present a
comprehensive analysis for prompt $\jpsi$ and $\psits$ produced at
the Tevatron and the LHC within NRQCD.  For prompt $\jpsi$, we have
taken feeddown contributions from $\chi_{cJ}(J=0,1,2)$ and $\psits$
into account. Short-distance coefficients for all important CO
Fock states are computed up to $\mathcal{O}(\alpha_S^4)$, i.e. at NLO
in $\alpha_S$. Color-singlet LDMEs of $\jpsi$, $\chi_{cJ}$ and
$\psits$ are estimated by using potential model~\cite{Eichten:1995ch},
while CO LDMEs are estimated by fitting experimental
data. For $\chi_{cJ}(J=0,1,2)$, there is only one independent CO LDME
$\mosochiz$. Its value can be fixed by fitting the Tevatron data
$\frac{\sigma(\chict\to\jpsi\gamma)}{\sigma(\chico\to\jpsi\gamma)}$~\cite{Abulencia:2007bra}
as done in ref.~\cite{Ma:2010vd}. For $\jpsi$ or $\psits$, there are
three independent CO LDMEs, i.e. $\mopajpsit$, $\mopbjpsit$ and $\mopcjpsit$.
From the decomposition of short-distance coefficients for $\pj$, we
understand that it is difficult to extract the three independent CO
LDMEs from the hadronic data even after including polarisation data.
What we can determine unambiguously is two linear combinations of
these three CO LDMEs. Their values were already extracted in
refs.~\cite{Ma:2010yw,Ma:2010jj} with $p_{T\rm{cut}}=7\rm{GeV}$. However, we still need three CO
LDMEs instead of two linear combinations to predict the yields and
polarizations for prompt $\jpsi$ and $\psits$ in various rapidity
regions. We assume all CO LDMEs are of positive signs, which are in
contrast to other groups' assumptions~\cite{Butenschoen:2012px,Gong:2012ug}. The result obtained under our
assumption is consistent with the observed relative magnitudes of
polarization in the forward rapidity interval and in the central rapidity
interval. Based on our assumption, we can provide more satisfactory predictions of both
yields and polarizations $\lambda_{\th}$ in the helicity frame for prompt
$\jpsi$, which is almost unpolarized at hadron
colliders. But we are unable to explain the
polarization of prompt $\psits$ based on the old fit in ref.~\cite{Ma:2010yw}. We thus checked the $\psits$ data and performed a new fit to the Tevatron data with $p_{T\rm{cut}}=11\rm{GeV}$, which gives a better description for the polarization data of $\psits$.

However, on the theoretical side, it is still needed to understand why we have to use such a large $p_T$ cutoff, which is much larger than the quarkonium mass. It might be possible that the NRQCD factorization formula may not be applicable if $p_T$ is not large enough, which were also pointed out by the authors in refs.~\cite{Bodwin:2014gia,Faccioli:2014cqa}. Recently, it was found that the $J/\psi$ production in small $p_T$ regime may be described by a CGC+NRQCD formalism \cite{Ma:2014mri}. In a moderate $p_T$ regime, say $p_T\sim 5-7$GeV, the CGC+NRQCD results match smoothly to our NLO NRQCD results \cite{Ma:2014mri}, and thus the $J/\psi$ production in the whole $p_T$ regime may be described. It will be interesting to see whether the $\psits$ production in small and moderate $p_T$ regime can be described in the same way.  In recent years, several other efforts are made by people to understand the quarkonium production mechanism, including the relativistic corrections \cite{Fan:2009zq,Xu:2012am}, the small $p_T$ regime resummation \cite{Sun:2012vc}, and the large $p_T$ regime factorization and resummation \cite{Kang:2011mg,Fleming:2012wy, Fleming:2013qu, Kang:2014tta, Kang:2014pya, Ma:2013yla,Ma:2014eja,Bodwin:2014gia,Ma:2014svb}. These works will provide more precise predictions for the quantitative understanding of quarkonium production.   Moreover, other quarkonium associated production processes (e.g. double $J/\psi$ production~\cite{Lansberg:2013qka,Sun:2014gca,Lansberg:2014swa}) and/or other observables (e.g. fragmenting jet functions~\cite{Baumgart:2014upa}) may also reveal the quarkonium production mechanism at the LHC in the future. On the experiment side, more precise measurements on the yields and especially the polarizations of heavy quarkonia are definitely needed to further clarify the present issues in  quarkonium production.

\begin{acknowledgments}
We thank D.~Price for useful discussions.
This work was supported in part by the National Natural Science
Foundation of China (No 11075002, No 11021092), and the Ministry of
Science and Technology of China (2009CB825200). Y.Q.Ma was supported
by the U.S. Department of Energy, under Contract No.
DE-AC02-98CH10886.

\end{acknowledgments}




\providecommand{\href}[2]{#2}\begingroup\raggedright\endgroup

\end{document}